\documentstyle[12pt,graphicx]{article}
\begin{document}
\makeatletter
\@addtoreset{equation}{section}
\makeatother
\renewcommand{\theequation}{\thesection.\arabic{equation}}
\baselineskip 15pt

\title{\bf Local measurements of nonlocal
observables and the relativistic reduction
process}

\author{GianCarlo Ghirardi\footnote{e-mail:
ghirardi@ts.infn.it}\\ {\small Department of
Theoretical Physics of the University of
Trieste, and}\\ {\small the Abdus Salam
International Centre for Theoretical Physics,
Trieste, Italy.}}

\date{}

\maketitle

\begin{abstract}
In this paper we reconsider the constraints which are imposed by relativistic
requirements to any model of dynamical reduction. We review the debate on
the subject
and we call attention on the fundamental contributions by Aharonov and
Albert. Having
done this we present a new formulation, which is
much simpler and more apt for our analysis, of the proposal put forward by
these authors to perform  measurements of nonlocal observables by means of
local
interactions and detections. We take into account recently proposed
relativistic
models of dynamical reduction and we show that, in spite of some mathematical
difficulties related to the appearence of divergences, they represent a
perfectly
appropriate conceptual framework which meets all necessary requirements for a
relativistic account of wave packet reduction. Subtle questions like the
appropriate
way to deal with counterfactual reasoning in a relativistic and nonlocal
context are
also analyzed in detail.
\end{abstract}

\section{Introduction}

As it is well known, standard quantum mechanics is characterized by irreducible
stochastic features which enter into play and give rise to puzzling
situations in connection with the measurement process or, more appropriately
and more generally, with the so called macro-objectification problem. It is
useful to stress that it is precisely with reference to such processes
which, when one takes further into account the unavoidable nonlocal
character of the theory, the problem of the causal relations between events
(such as the measurement outcome in one region and the transition from
potential to actual physical situations in another) emerges as a central
one. It goes without saying that an adequate discussion of such questions
requires a relativistic approach. Our main concern here will be the analysis
of statevector reduction in a relativistic quantum context and the
identification
of the basic features which any relativistic reduction mechanism must
exhibit. In
particular we will show that recent attempts in this direction [1]-[6],
even though not fully satisfactory due to some technical problems (such as the
appearence of untractable divergencies) have given clear indications about
the line
of thought one should follow to account for the nonlinear and stochastic
reduction
process meeting the strict demands of special relativity\footnote{Some of the
arguments we will discuss in this paper have been already presented in
refs.\cite{G1} and \cite{G2}. However, the new approach to the problem of local
measurements of nonlocal observables which is presented in section 3 will
allow us to develope in a clearer and more convincing way  our
arguments  and to draw the relevant conclusions of the final section.}.

In section 2 we
reconsider the crucial aspects of the problem under investigation by
following the
recent lucid presentation of the matter by Breuer and Petruccione
\cite{BrPe}. Particular attention will be devoted to the fundamental
contributions
to the subject by Aharonov and Albert \cite{AhAl}, \cite{AhAl2} and
\cite{AhAl3}.
These authors have identified the necessary formal features that any acceptable
relativistic reduction mechanism must exhibit and have shown that all previous
attempts in this direction were fundamentally unsatisfactory. The key argument
which has led Aharonov and Albert to draw such a conclusion is their proof
\cite{AhAl2} that one can measure nonlocal observables by resorting to local
interactions between appropriately (and smartly) chosen quantum systems,
followed
by local detection processes.

There are, however, two reasons which require to deepen and to improve
the line of thought of these authors. First, their explicit example of a
local measurement of nonlocal observables is quite formal and rests on the
consideration of nonnormalizable states. Secondly, in spite of the fact
that they have given clear indications about the general formal features
of any acceptable relativistic dynamical reduction mechanism, they have
made no attempt to work out an explicit example of such a dynamics. On the
other hand, as already mentioned, in recent years precise relativistic
models of
dynamical reduction have been formulated [1]-[6]. It is one of the aims of this
paper to investiagte critically such models from the point of view of the
analysis
of Aharonov and Albert.

To this purpose it is useful to work out, first of all, a remarkably simplified
version of the measurement procedure suggested by the above authors in
ref.\cite{AhAl2}.
This is achieved in section 3 by relating the relevant physical process to
degrees of
freedom whose associated Hilbert space is finite--dimensional. In this way
the treatment
turns out to be extremely simple and intuitive and the procedure easy to
implement in
practice. This simplified version of the proposal of ref.\cite{AhAl2} will
allow us to
focus the essential features of any relativistic reduction theory and to
show that such
features are essential and natural ingredients of the theoretical framework
presented in
refs.\cite{P}-\cite{G2}.   Accordingly, in the rest of the paper we will
discuss how
such precise theoretical schemes account for the relativistic
macro-objectification
process and we will show that they lead to a perfectly coherent view which
allows one
to analyze basic issues such as the\textit{\ attribution of objective
properties} to
individual physical systems, the necessary generalization of the concept of
\textit{event} and the appropriate way to resort to \textit{counterfactual
arguments}
within a relativistic and nonlocal context.

\section{Relativistic reduction processes}

This section is devoted to reviewing some of the issues of the central theme
of the paper, i.e., the relativistic aspects of the reduction process.

\subsection{Local relativistic reduction processes}

The main problems which one meets in trying to generalize the
nonrelativistic process of statevector reduction derive from the assumed
instantaneity of such a process. As lucidly described in ref.\cite{BrPe} one
can consider the case in which one has a system whose associated
wavefunction has an appreciable spatial extension and, at time $t=0,$ is
found at $x=0$ by a detector which is placed there. The problem one has to
face is quite obvious: even if one were able to account for the local
position measurement in terms of a local covariant interaction between the
measured object and the measuring device, the ensuing picture would
obviously turn out not to be covariant for the very simple reason that in
any other reference frame the space-like surface $t=0$ is not an equal time
surface; consequently, the reduction cannot be instantaneous for any
observer in motion with respect to the original one.

Hellwig and Kraus \cite{HeKr} have proposed to circumvent the above
difficulty by postulating that in a local measurement process statevector
reduction takes place along the past light cone originating from the
space-time point at which the covariant system-apparatus interaction occurs.
In contrast with the case mentioned above it is obvious that the
proposed prescription is manifestly covariant since the light cone from a given
point is the same in all reference frames. These authors have also shown that
their prescription leads to the correct quantum predictions concerning the
probabilities of the outcomes of local measurements of local observables. In
the early stages of the debate about relativistic reduction processes the
expression ``local'' has been used in a quite loose way which we will follow
in this subsection with the purpose of making quite intuitive the problems
one meets. Suppose our system, let us say an elementary particle, prior to
the local position measurement at $t=0,$ $x=0,$ is described, in a given
reference
frame, by a wavefunction which has an appreciable extension in space (the
limiting
case would be the one of a system in a state of definite momentum). A
measurement
in which the particle is found at $x=0$ will, according to the Hellwig and
Kraus prescription, lead to a statevector which is different from zero only
within the past light cone from the origin, contradicting the fact that at a
time prior to $t=0$ the wavefunction extends over a much larger region as,
e.g.,
in the limiting case of a momentum
eigenstate, which, if subjected to a momentum measurement, would remain
unaltered. According to Hellwig and Kraus the way out
from such a difficulty derives from assuming that, e.g., the alleged momentum
measurement cannot be performed by resorting to local interactions. Actually,
already in 1931, Landau and Peierls \cite{LaPe} had suggested that all nonlocal
quantities, like the momentum operators, cannot be considered as observables in
relativistic quantum theories. Hellwig and Kraus adopt this point of wiev and,
consequently, they dismiss as non pertinent any criticism concerning their
proposal. Here comes the fundamental contribution by Aharonov and Albert
\cite{AhAl2}. They have shown that it is actually possible to measure nonlocal
observables and, even more important, that this can be done by local
interactions
and local detections.

Before going on, it is appropriate to stress a point of great conceptual
relevance for the subject of this paper. In the debate we have just reviewed
it was always assumed (more or less tacitly) that the wavefunction $\Psi
(x,t),$ in the relativistic case must be considered as a \textit{function on
the space-time continuum}. As we will see, the analysis of Aharonov and
Albert requires a radical change of perspective about this fact.

\subsection{Nonlocal measurements in a relativistic context and their
difficulties}

As already anticipated, within a relativistic context, nonlocal observables
raise
particular problems in connection with measurement processes. To
better focus this point, in place of the rather vague considerations of the
previous
subsection related to momentum measurements, we present a slightly modified
version
of a simple and nice example by Breuer and Petruccione \cite{BrPe}.
\begin{figure}[htb]
\begin{center}
{\includegraphics[scale=0.5]{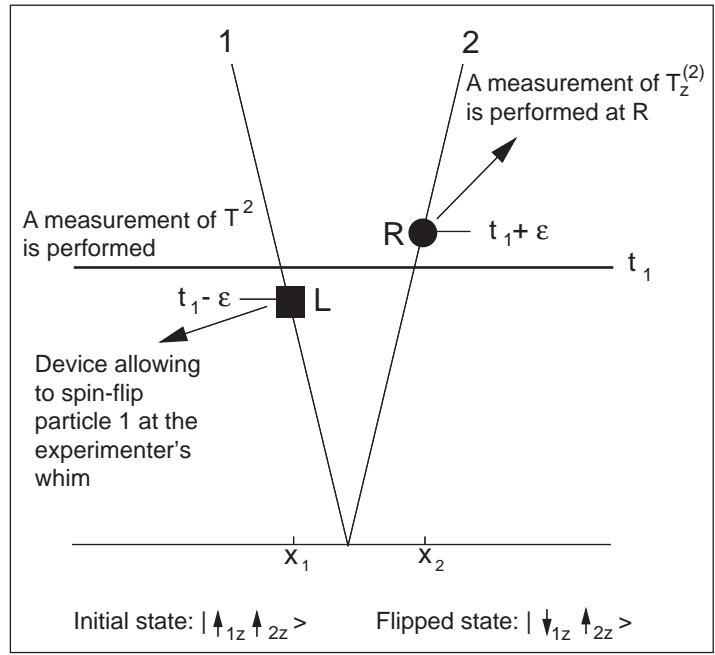}}
\caption{An ideal experiment which points out the restrictions on nonlocal
measurements imposed by causality}
\end{center}
\end{figure}
Let us consider
(Fig.1) a given inertial reference frame and a pair of particles having
space plus internal degrees of freedom. For simplicity, since these
assuptions do not change in any way whatsoever the conceptually relevant
aspects of the analysis, we will disregard the spreading of the wavepackets
of the two particles and we will assume that the internal degrees of
freedom refer to quantities, like the isospin, which do not transform under
Lorentz transformations. Suppose the two particles are created, e.g., in a
decay process at a given objective space-time point and that they propagate
in opposite directions. Concerning the internal degrees of freedom, let us
assume that the two particles have isospin $1/2$ and are in the common
eigenstate of $T^{2}$ and $T_{z}$ belonging to the eigenvalues (in the
appropriate
units) $2$ and
$1$, respectively, i.e., in the state we will denote (in analogy with the
case of
spin variables) as $|\uparrow _{1z},\uparrow _{2z}>.$ Suppose that, along
the world line of particle 1 (the one propagating at left), at a precise
space-time point $L(x_{1},t_{1}-\epsilon )$, there is a device which can be
switched
on and off at the experimenter's whim, whose only (local) effect is that of
inducing
a isospin flip of the particle interacting with it when it is on, while no
change occurs when it is off. Suppose also that, at time $t_{1}$, a
measurement of the nonlocal observable $T^{2}$ is performed on the composite
system. We stress that since the two particles are in different space
regions the measurement of $T^{2}$ is unavoidably nonlocal. Finally let us
assume that at time $t_{1}+\epsilon ,$ at a point $R(x_{2},t_{1}+\epsilon
)$ on the
world line of particle 2 there is an apparatus measuring $T_{2z},$ the $z$
component
of the isospin of particle 2.

Given these premises one can argue as follows:

\begin{itemize}
\item Suppose the apparatus at $L$ is off. Then the state of the composite
system at $t_{1}$ is $|\uparrow _{1z},\uparrow _{2z}>,$ which,
being an eigenstate of $T^{2},$ is left unchanged by the measurement of this
observable. There follows that in the subsequent measurement of
$T_{2z}$ the outcome +1 is obtained with certainity.

\item Suppose now the apparatus at $L$ is on. Then the state of the composite
system at $t_{1}-\epsilon$ becomes $|\downarrow _{1z},\uparrow _{2z}>,$
which is
the superposition with equal squared amplitudes of the eigenstates of $T^{2}$
belonging to the eigenvalues 0 and 2, respectively. The measurement taking
place at $t_{1}$ will then reduce the state to one of such eigenstates,
i.e., $\frac{1}{\sqrt{2}}[|\downarrow _{1z},\uparrow _{2z}>\mp $ $|\uparrow
_{1z},\downarrow _{2z}>],$ which assign probability $1/2$ both to the
outcome +1 and to --1 in the measurement of $T_{2z}$ at $t_{1}+\epsilon $.
If the
observer at $R$ is informed about the experimental set-up, in the cases in
which he gets the outcome --1 he can infer that the observer at $L$ has
actually decided to induce the spin flip of particle 1. This argument shows
quite nicely that nonlocal measurements can lead to a violation of causality
allowing superluminal communication and makes clear while such measurements
must be excluded if one wants to take the position of Hellwig and Kraus
about statevector reduction.
\end{itemize}

\subsection{The Aharonov and Albert procedure}

The simplest example of a measurement of a nonlocal observable proposed by
these authors is, in our language, a measurement of the $z-$component of the
total isospin  $T_{z}=T_{1z}+T_{2z}$ of the previous system. The smart trick
consists in considering an apparatus  constituted by two subsystems (probes)
whose world lines intersect the world lines of particles 1 and 2 (once more we
disregard the spreading associated with the free motion of the probes) and
considering ``internal generalized coordinates $\widehat{q}_{1}$ and $%
\widehat{q}_{2}$''of the probes. The local interactions of the probes with
the particles are described by the hamiltonian:
\begin{equation}
H_{I}=g(t)\left[ \widehat{q}_{1}T_{1z}+\widehat{q}_{2}T_{2z}\right] ,
\label{e1}
\end{equation}
where $g(t)$ is a time dependent coupling constant vanishing outside a small
time interval $(t_{1},t_{2})$. Finally, one assumes that the probes are, at
any time $t_{0}$ preceeding their interactions with the particles, in an
entangled state of the kind of the one considered in the celebrated EPR
paper, i.e., one in which both the coordinates $\widehat{q}_{1}$ and $%
\widehat{q}_{2}$ and the conjugated momenta $\widehat{\pi }_{1}$ and $%
\widehat{\pi }_{2}$ are perfectly correlated:
\[
\widehat{\pi }_{1}(t_{0})+\widehat{\pi }_{2}(t_{0})=0
\]
\begin{equation}
\widehat{q}_{1}(t_{0})-\widehat{q}_{2}(t_{0})=0.  \label{e2}
\end{equation}
As in the standard measurement scheme by von Neumann the local interactions
described by eq.(\ref{e1}) imply the following equation of motion for the
total canonical momentum $\pi (t)$:
\begin{equation}
\frac{\partial \pi (t)}{\partial t}=i[H_{I},\pi (t)]=-g(t)\left[
T_{z1}+T_{z2}\right] =-g(t)T_{z,}  \label{e3}
\end{equation}
so that one can infer the value of $T_{z}$ from the change of the total
momentum:
\begin{equation}
T_{z}=-\frac{\pi (t>t_{2})}{\int_{t_{1}}^{t_{2}}dtg(t)},  \label{4}
\end{equation}
where we have taken into account that $\pi (t<t_{1})=0,$ as shown by the
first of eqs.(\ref{e2}). To make eq.(\ref{4}) more understandable to the
reader, we remark that the local interaction of one probe with the
corresponding particle decreases (increases) the value of the associated
momentum according whether the particle is in the eigenstate $T_{iz}=+1$ $%
(-1),$ respectively. Accordingly, the total momentum changes when the two
particles have parallel $z-$isospin components and remains unchanged when
their projections are opposite.

It is extremely important to realize that the measurement is a genuinely
nonlocal measurement performed by means of local interactions and
detections. In fact it is easy to convince oneselves that the knowledge,
after the measurement, of the value of the momentum of only one of the
probes does not allow to draw any conclusion  referring to the
isospin component of the particle which has interacted with the probe, or
to the component of the total isospin. This point will become much more
clear when
we will consider the simplified version of a measurement of this kind which
is the
subject of the following Section.

Some remarks are at order:

\begin{itemize}
\item The two probes considered in refs.\cite{AhAl2} have translational
degrees of freedom related to their propagation and interactions with the
particles
and further internal degrees of freedom which, however, are associated to
continuous generalized coordinates and momenta. Moreover, the state of the
probes
before the measurement exhibits perfect ``internal" momentum and position
correlations, just as the state of the original EPR argument. Leaving aside the
problem of the difficulty of preparing such an entangled state in these
continuous
variables, the resulting state turns out to be nonnormalizable. It
should be clear to the reader that the very cute proposal by Aharonov and
Albert corresponds to a quite idealized situation. Just in the same way in
which Bohm has felt the necessity of rephrasing the EPR argument resorting
to the
consideration of spin degrees of freedom, transforming it from a gedanken to a
feasible experiment, we consider it useful to perform an analogous step with
reference to the process under discussion.

\item The above procedure has been generalized by the authors of refs.%
\cite{AhAl2} to measure the square $T^{2}$ of the total isospin of the
composite system. Such a step involves the use of three pairs of entangled
probes which interact with particles 1 and 2. Only the comparison of the
outcomes obtained at the two wings of the apparatuses devised to measure the
momenta of the three probes allows, in the case in which all pairs of
momenta sum up to zero, to assert that the system has been found in the
singlet state. The measurement is fundamentally a quantum nondemolition
measurement for the singlet state, while it is not a measurement of the
standard type for the states of the three dimensional manifold $T^{2}=2.$
Once more these subtle questions will become clear in the modified version
of a local measurementof nonlocal observables we are going to discuss.
\end{itemize}

\section{A new procedure for a local measurement of nonlocal observables}

In this Section we present a possible experimental setup for the measurement
of the observables $T_{z}$ and $T^{2}$ of the system \textit{S} of two
particles of isospin 1/2
which is based on local interactions between the particles and probe
particles whose internal degrees of freedom are accounted by the states of a
three-dimensional Hilbert space (Fig.2).
\begin{figure}[htb]
\begin{center}
{\includegraphics[scale=0.5]{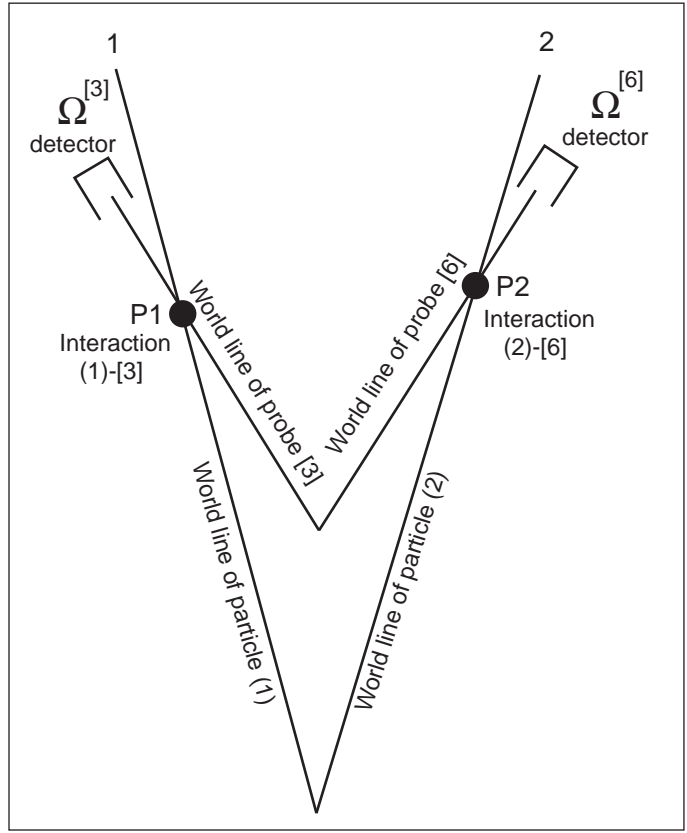}}
\caption{The way to perform a
nonlocal measurement by resorting to local interactions and detections.}
\end{center}
\end{figure}
To begin, we consider, as before,
two particles 1 and 2 which are produced at a given space-time point and
propagate along two world lines. Since we disregard the spreading of the
wavefunctions of the particles, their quantum features are entirely
described by the vectors of the four-dimensional Hilbert spaced spanned by
the states (with obvious meaning of the symbols):
\begin{equation}
|\uparrow _{1z},\uparrow _{2z}>,\qquad |\uparrow _{1z},\downarrow
_{2z}>,\qquad |\downarrow _{1z},\uparrow _{2z}>,\qquad |\downarrow
_{1z},\downarrow _{2z}>.  \label{e5}
\end{equation}
We consider also a system $\Sigma $ of two probe particles (which for
reasons which will become clear in what follows we identify as 3 and 6)
whose world lines intersect, at the space time points $P1$ and $P2$, the world
lines of the system particles. The probe particles, as already stated,
besides their configurational degrees of freedom (which we treat once more as
classical), have internal degrees of freedom associated to a
three-dimensional Hilbert space. We denote as
\begin{equation}
|-1>_{i},\qquad |0>_{i},\qquad |+1>_{i},\qquad (i=3,6),  \label{e6}
\end{equation}
the complete set of the eigenstates of an observable $\Omega ^{(i)}$ of the
Hilbert space of the i-th probe and by
\begin{equation}
|a,b>_{3,6}\equiv |a>_{3}\otimes |b>_{6},\qquad (a,b=-1,0,+1)  \label{e7}
\end{equation}
the elements of the corresponding basis in the nine-dimensional Hilbert
space of $\Sigma $.
The initial state of the probe is assumed to be the normalized entangled
state of particles 3 and 6:
\begin{equation}
|\Phi >_{3,6}=\frac{1}{\sqrt{3}}\left[
|0,0>_{3,6}+|+1,-1>_{3,6}+|-1,+1>_{3,6}\right] .  \label{e8}
\end{equation}

The system and probe particles interact locally (constituent 1 (2) of $S$ with
constituent 3 (6) of $\Sigma $) for a given
time interval, the effect of the interaction being described by the unitary
operator $U_{z}:$%
\begin{equation}
U_{z}=\left[ P_{z+}^{(1)}P_{L}^{(3)}+P_{z-}^{(1)}P_{R}^{(3)}\right] \otimes
\left[ P_{z+}^{(2)}P_{L}^{(6)}+P_{z-}^{(2)}P_{R}^{(6)}\right] ,  \label{e9}
\end{equation}
where the operators $P^{(i)}_{z\pm}$ are
the projection operators on the eigenmanifolds of
$T_{iz}$:
\[
P_{z+}^{(i)}=|\uparrow _{iz}><\uparrow _{iz}|,\qquad
P_{z-}^{(i)}=|\downarrow _{iz}><\downarrow _{iz}|,\qquad (i=1,2)
\]
and the operators $P^{(j)}_{R,L}$ act on the probe space and have,
 in the basis (%
\ref{e6}), the
representation indicated below:
\begin{equation}
P_{L}^{(j)}=\left(
\begin{array}{lll}
0 & 0 & 1 \\
1 & 0 & 0 \\
0 & 1 & 0
\end{array}
\right) _{j},\qquad P_{R}^{(j)}=\left(
\begin{array}{lll}
0 & 1 & 0 \\
0 & 0 & 1 \\
1 & 0 & 0
\end{array}
\right) _{j},\qquad (j=3,6).  \label{e10}
\end{equation}
We note that the operators $P_{L}^{(j)}$ and $P_{R}^{(j)}$ act in the
following way on the states of the base of probe j:
\begin{equation}
P_{L}^{(j)}|-1>_{j}=|+1>_{j},\qquad P_{L}^{(j)}|0>_{j}=|-1>_{j},\qquad
P_{L}^{(j)}|+1>_{j}=|0>_{j},
\end{equation}
\begin{equation}
P_{R}^{(j)}|-1>_{j}=|0>_{j},\qquad P_{R}^{(j)}|0>_{j}=|+1>_{j}\qquad
P_{R}^{(j)}|+1>_{j}=|-1>_{j},  \label{e11}
\end{equation}
so that they produce an anticlockwise and a clockwise permutation of the
three states ($ |-1>, |0>, |+1> $) of the basis, respectively. We also note the
following properties of these operators:
\begin{equation}
P_{R}^{(j)\dagger }=P_{L}^{(j)},\qquad P_{L}^{(j)\dagger
}=P_{R}^{(j)},\qquad P_{R}^{(j)}P_{L}^{(j)}=P_{L}^{(j)}P_{R}^{(j)}=1^{(j)}.
\label{e12}
\end{equation}
The above relations imply:
\begin{eqnarray}
U_{z}^{\dagger }U_{z} &=&\left[
P_{z+}^{(1)}P_{R}^{(3)}+P_{z-}^{(1)}P_{L}^{(3)}\right] \otimes \left[
P_{z+}^{(2)}P_{R}^{(6)}+P_{z-}^{(2)}P_{L}^{(6)}\right] \cdot   \label{e13}
\\
&&\left[ P_{z+}^{(1)}P_{L}^{(3)}+P_{z-}^{(1)}P_{R}^{(3)}\right] \otimes
\left[ P_{z+}^{(2)}P_{L}^{(6)}+P_{z-}^{(2)}P_{R}^{(6)}\right]   \nonumber \\
&=&\left[ P_{z+}^{(1)}1^{(3)}+P_{z-}^{(1)}1^{(3)}\right] \left[
P_{z+}^{(2)}1^{(6)}+P_{z-}^{(2)}1^{(6)}\right]
=1^{(1)}1^{(2)}1^{(3)}1^{(6)}=1,  \nonumber
\end{eqnarray}
so that $U_{z}$ is a unitary operator.

Given the above equations it is immediate to evaluate the effect of applying
$U_{z}$ to the direct product of any one of the states (\ref{e5}) times the
initial state of the probe $|\Phi >_{3,6}.$ To this purpose it is useful to
introduce the following normalized states of the probe system (with obvious
meaning of the symbols):
\begin{eqnarray}
|\Pi (1,-2) >_{3,6}&=&\frac{1}{\sqrt{3}}\left[
|-1,-1>_{3,6}+|0,+1>_{3,6}+|+1,0>_{3,6}\right]   \label{e14} \\
|\Pi (2,-1) >_{3,6}&=&\frac{1}{\sqrt{3}}\left[
|+1,+1>_{3,6}+|0,-1>_{3,6}+|-1,0>_{3,6}\right] .  \nonumber
\end{eqnarray}
As easily checked we then have:
\begin{eqnarray}
U_{z}\left[ |\uparrow _{1z},\uparrow _{2z}>\otimes |\Phi >_{3,6}\right]
&=&|\uparrow _{1z},\uparrow _{2z}>\otimes |\Pi (1,-2)>_{3,6}  \label{e16} \\
U_{z}\left[ |\downarrow _{1z},\downarrow _{2z}>\otimes |\Phi >_{3,6}\right]
&=&|\downarrow _{1z},\downarrow _{2z}>\otimes |\Pi (2,-1)>_{3,6}  \nonumber
\\
U_{z}\left[ |\uparrow _{1z},\downarrow _{2z}>\otimes |\Phi
>_{3,6}\right]  &=&|\uparrow _{1z},\downarrow _{2z}>\otimes |\Phi >_{3,6}
\nonumber \\
U_{z}\left[ |\downarrow _{1z},\uparrow _{2z}>\otimes |\Phi
>_{3,6}\right]  &=&|\downarrow _{1z},\uparrow _{2z}>\otimes |\Phi >_{3,6}
\nonumber
\end{eqnarray}
The model we have presented, in complete analogy with the one proposed by
Aharonov and Albert, describes a nonlocal measurement procedure (in terms of
local interactions and detections) of the nonlocal observable $T_{z},$ the
\textit{z}-component of the total isospin of the composite system $S$. In
fact, in
order to complete the measurement, we have only to put on the world lines
of the
probe particles two apparata measuring the observables
$\Omega^{(i)}$(\textit{i}=3,6),
to register the obtained outcomes and to sum them. If they sum up to 0,
then reduction
takes place to a state for which $T_{z}=0,$ if they sum up to either 1 or
--2 then
reduction takes place to the state for which $T_{z}=1,$ finally, if they
sum up to
either 2 or --1, reduction takes place to the state for which $ T_{z}=-1.$

It is of great relevance to remark that for each of the states of the probes
appearing at the right hand side of eq.(\ref{e16}) all possible eigenvalues (%
$\omega _{k}^{(i)}=-1,0,1)$ of the observable $\Omega ^{(i)}$ referring to
each probe particle appear with equal weights. Thus, knowledge of the
outcome, e.g., of the detector at the right wing of the apparatus, does not
give any information whatsoever about the isospin component of the particle
with which the probe particle has interacted locally, nor about the
component of the total isospin. In this precise sense the measurement is
genuinely nonlocal (as it must be, since the two particles lie in far apart
regions).

We can now proceed to generalize the previous procedure to a measurement
process for $T^{2}.$ To this purpose we have to resort to a more complicated
device corresponding to the production of three entangled pairs of probe
particles which interact locally with the particles whose total isospin is
being measured. So, besides the pair (3-6) of probe particles we will
consider also the pairs\footnote{%
We have decided to use the indices (3,6), (2*,4*) and (3*,6*) to refer to
the three pairs of probe particle for obvious reasons: the specification 3
and 6 (2 and 4) refer to degrees of freedom which are locally coupled to the
third, i.e. the \textit{z}, (the second, i.e. the \textit{y}) component of
the isospin of the measured particles. The stars are used to avoid confusion
with the indices 1,2 of the measured particles or to stress that the last
pair (3*,6*) of probe particles is different from the first one. } (2*-4*)
and (3*-6*). The initial state of the whole probe will simply be the direct
product of three states like (\ref{e8}):
\begin{equation}
|\Phi >=|\Phi >_{3*,6*}\otimes |\Phi >_{2*,4*}\otimes |\Phi >_{3,6}
\label{e17}
\end{equation}
and the unitary evolution operator $U$ for the whole system is assumed to
be the
product of three operators: $\widetilde{U}_{z}U_{y}U_{z}$. In turn, $U_{z}$
is the
operator defined in (\ref{e9}), $U_{y}$ has exactly the same form of $U_{z}$
with the operators
$P_{y+}^{(1)},P_{y-}^{(1)},P_{y+}^{(2)},P_{y-}^{(2)}$  replacing the
corresponding
ones with the index  $z$, and the
operators $P_{L}^{(2*)},P_{R}^{(2*)},P_{L}^{(4*)},P_{R}^{(4*)}$
 replacing the corresponding ones with the  indices  $3$ and $6$.
Finally $\widetilde{U}_{z}$ coincides with $U_{z}$ with the replacement of
the indices 3 and 6 by the corresponding indices 3* and 6*.

To see how the mechanism works we have, first of all, to express the states (%
\ref{e5}) in terms of the corresponding states of the type $|\uparrow
_{1y},\uparrow _{2y}>$ etc., in order to evaluate the effetct of applying
the operator $U_{y}.$ Moreover, since now we are interested in identifying
the eigenstates of $T^{2}$, it is useful to replace the states $|\uparrow
_{1z,}\downarrow _{2z}>$ and $|\downarrow _{1z,}\uparrow _{2z}>$ by their
simmetrical and skew-symmetrical combinations:
\begin{eqnarray}
|Triplet >_{z} &=& \frac{1}{\sqrt{2}}\left[ |\uparrow _{1z},\downarrow
_{2z}>+|\downarrow _{1z},\uparrow _{2z}>\right] ,  \label{e15} \\
|Singlet >_{z} &=&\frac{1}{\sqrt{2}}\left[ |\uparrow _{1z},\downarrow
_{2z}>-|\downarrow _{1z},\uparrow _{2z}>\right]   \nonumber
\end{eqnarray}
and analogous ones for the states\footnote{%
We remind that while no specification is necessary for the singlet state, in
the case of the triplet state the specifications \textit{z} or\textit{\ y}
are essential since they are genuinely different states.} $|\uparrow
_{1y,}\downarrow _{2y}>$ and $|\downarrow _{1y,}\uparrow _{2y}>$.

With these premises an elementary (but tedious) calculation shows the effect
of applying the unitary operator $U$ to the isospin singlet and triplet
states:
\begin{equation}
U|Singlet>\otimes |\Phi >=|Singlet>\otimes |\Phi >
\end{equation}

\begin{eqnarray}
\lefteqn{U| \uparrow_{1z},\uparrow _{2z}>\otimes |\Phi >=}  \\
& & +\left\{ \frac{1}{4}|\Pi (1,-2) >_{3*,6*}\otimes \left[ |\Pi %
(1,-2)>_{2*,4*}+|\Pi (2,-1)>_{2*,4*}+2|\Phi >_{2*,4*}\right] \otimes
|\uparrow _{1z},\uparrow _{2z}> \right. \nonumber \\
& & -\frac{1}{4}|\Pi (2,-1) >_{3*,6*}\otimes \left[ |\Pi (1,-2)>_{2*,4*}+|\Pi %
(2,-1)>_{2*,4*}-2|\Phi >_{2*,4*}\right] \otimes |\downarrow _{1z},\downarrow
_{2z}> \nonumber \\
& & \left. +\frac{i}{2\sqrt{2}}|\Phi >_{3*,6*}\otimes \left[ |\Pi
(1,-2)>_{2*,4*}-|%
\Pi (2,-1)>_{2*,4*}\right] |Triplet>_{z}\right\}\otimes |\Pi (1,-2)>_{3,6};
\nonumber \end{eqnarray}

\begin{eqnarray}
\lefteqn{U| \downarrow_{1z},\downarrow _{2z}>\otimes |\Phi >= }\\
& & \left\{  -\frac{1}{4}|\Pi (1,-2) >_{3*,6*}\otimes \left[ |%
\Pi (1,-2)>_{2*,4*}+|\Pi (2,-1)>_{2*,4*}-2|\Phi >_{2*,4*}\right] \otimes
|\uparrow _{1z},\uparrow _{2z}> \right. \nonumber \\
& & +\frac{1}{4}|\Pi (2,-1) >_{3*,6*}\otimes \left[ |\Pi (1,-2)>_{2*,4*}+|\Pi %
(2,-1)>_{2*,4*}+2|\Phi >_{2*,4*}\right] \otimes |\downarrow _{1z},\downarrow
_{2z}> \nonumber \\
& & \left. -\frac{i}{2\sqrt{2}}|\Phi  >_{3*,6*}\otimes \left[ |\Pi
(1,-2)>_{2*,4*}-|%
\Pi (2,-1)>_{2*,4*}\right] |Triplet>_{z}\right\}\otimes |\Pi (2,-1)>_{3,6};
\nonumber \end{eqnarray}

\begin{eqnarray}
\lefteqn{U|Triplet >_{z}\otimes |\Phi >=} \\
& & \left\{ -\frac{i}{2\sqrt{2}}|\Pi (1,-2) >_{3*,6*}\otimes \left[ |\Pi %
(1,-2)>_{2*,4*}-|\Pi (2,-1)>_{2*,4*}\right] \otimes |\uparrow _{1z},\uparrow
_{2z}> \right. \nonumber \\
& & +\frac{i}{2\sqrt{2}}|\Pi (2,-1) >_{3*,6*}\otimes \left[ |\Pi %
(1,-2)>_{2*,4*}-|\Pi (2,-1)>_{2*,4*}\right] \otimes |\downarrow
_{1z},\downarrow _{2z}> \nonumber \\
& & \left. +\frac{1}{2}|\Phi >_{3*,6*}\otimes \left[ |\Pi
(1,-2)>_{2*,4*}+|\Pi %
(2,-1)>_{2*,4*}\right] \otimes |Triplet>_{z}\right\}\otimes |\Phi >_{3,6}.
\nonumber \end{eqnarray}

Suppose now that there are six local detectors (three at each wing) devised to
measure the observables $\Omega ^{(s)},$  $s=$ 3,6,2*,4*,3*,6* and denote as
$\omega ^{(s)}$ the obtained outcomes. The above relations exhibit some
interesting features:

i). If $\omega ^{(3)}+\omega ^{(6)}=\omega ^{(2*)}+\omega ^{(4*)}=\omega
^{(3*)}+\omega ^{(6*)}=0,$ then reduction has taken place to $|Singlet>.$

ii). If at least one of the  relations under i) is not satisfied, reduction has
taken place to a state belonging to the three-dimensional $T^{2}=2$
eigenmanifold.

This second case can be further analyzed according to the following table:
\[
\begin{array}{clclcl}
\makebox{If} &
\omega ^{(3*)}+\omega ^{(6*)} & =& \makebox{$1$ or  $-2$} & \makebox{reduction
has taken place to} & |\uparrow _{1z},\uparrow_{1z}> \\
\makebox{If} &
\omega ^{(3*)}+\omega ^{(6*)} & =& \makebox{$2$ or  $-1$} & \makebox{reduction
has taken place to} & |\downarrow _{1z},\downarrow_{1z}> \\
\makebox{If} &
\omega ^{(3*)}+\omega ^{(6*)} & =& \makebox{$0$ and $\omega ^{(2*)}+\omega
^{(4*)}\neq 0$}
& \makebox{reduction has taken place to} & |Triplet>_{z}.
\end{array}
\]

We can consider now an arbitrary state of the system, i.e. a linear
superposition of the four states (\ref{e5}),
\begin{equation}
|\Psi >=\alpha |\uparrow _{1z},\uparrow _{2z}>+\beta |\uparrow
_{1z},\downarrow _{2z}>+\gamma |\downarrow _{1z},\uparrow _{2z}>+\delta
|\downarrow _{1z},\downarrow _{2z}>  \label{e18}
\end{equation}
and use it to trigger our measuring devices. The resulting state $U|\Psi
>\otimes |\Phi >$ will be the linear combination with the same coefficients
of the states at the right hand side of eqs.(3.15-3.18). Let us write, for
simplicity such a state as:
\begin{equation}
U|\Psi >\otimes |\Phi >=\eta |Singlet>\otimes |\Phi >+|\Gamma >.
\label{e19}
\end{equation}
As already remarked, if the detection procedure of the probe particles
indicates that
$\omega ^{(3)}+\omega ^{(6)}=\omega ^{(2*)}+\omega ^{(4*)}=\omega
^{(3*)}+\omega
^{(6*)}=0$ (and this occurs with probability $|\eta |^{2}$) then reduction
takes place to the state $|Singlet>.$ In all other cases, even though
reduction leads
to a state of the eigemanifold $T^{2}=2,$ the measurement process  does not
respect the request of being ``moral'', i.e., the condition that the
reduced state be
the (normalized) projection of the state prior to the measurement on the
considered
eigenmanifold. Said differently, the measurement, when it yields the outcome
$T^{2}=2,$ turns out to be distorting for $T_{z}$. This is immediately seen by
choosing in (\ref{e18})
$\beta =\gamma =\delta =0$ and by looking at eq.(3.16) which shows that
there is a nonzero probability that the measurement leads to reduction on
one of the states  $|\downarrow _{z1},\downarrow _{z2}>$ and $|Triplet>_{z}.$
This is naturally related to the fact that, in a loose sense, one could
state that in order to measure $T^{2}$ one has to measure simultaneously
incompatible observables, such as $T_{1z}$ and $%
T_{1x,}$  of one of the particles. But for our purposes this nonideality of the
measurement for the states of the eigenmanifold $T^{2}=2,$ is not relevant.

Actually, the feature of the measurement process we have just analyzed, besides
being unavoidable, is an extremely positive one. In fact, if consideration
is given
to the situation discussed in Subsection 2.2, one easily sees that the
distorting
nature of the measurement eliminates the possibility of faster than light
effects.
From eq.(3.16) one sees that if a measurement of $T^{2}$ is performed on the
state $|\uparrow _{1z},\uparrow _{2z}>$, the probability of getting in a
subsequent measurement the outcome $T_{2z}=+1$ is no more equal to 1, but
precisely
to $1/2$. This agrees with the probability of getting such an outcome if one
performs the measurement of  $T^{2}$ on the state which has been subjected to a
spin flip of particle 1, i.e. the state $|\downarrow _{1z},\uparrow _{2z}>$, as
one derives immediately from a combined use of eqs.(3.15) and (3.18). On the
contrary, eqs.(3.12) show that a local measurement of the nonlocal observable
$T_{z}$ does not alter the fact that the probability of getting the
outcome $T_{2z}=+1$ remains unaltered (and equal to 1) when the spin of
particle 1
is flipped.

It is time to come to discuss the conceptual implications of the
actual possibility of  measuring nonlocal observables identified
by Aharonov and Albert and reformulated by us in this Section.

\section{The implications of the request of a covariant reduction}

The analysis of ref.\cite{AhAl2} and of the previous section rules out the
Hellwig and Kraus proposal for a covariant reduction mechanism and requires
to consider an alternative approach to the problem. This has been suggested
in another important paper \cite{AhAl3} by Aharonov and Albert. The
proposal  is quite
simple and natural, but it has far reaching consequences. One could
describe it by
stating that, in a certain sense,  they assume that reduction takes place
instantaneously in all inertial frames. To be more clear, let us consider an
objective space-time point $P$ where a local measurement occurs and the set
of all
space-like hyperplanes%
\footnote{%
Actually the above mentioned authors have considered more generally and more
appropriately arbitrary space like surfaces through $P$ and have related
each of them to the inertial frame in which the normal to the surface at the
considered point coincides with the time axis of the frame. This choice is
not only useful but necessary in a perspective like the one of dynamical
reduction models in which the dynamical equations, given the statevector on
an initial space-like surface, determine (stochastically) the statevector to
be attached to any space-like surface lying entirely in the future of the
initial one. Here, however, to allow the reader to grasp intuitively the
meaning of the proposal, we will resort to the consideration of space-like
hyperplanes.} through $P.$ Each such hyperplane is a $t=const$ surface for
an appropriate observer. The proposal of ref.\cite{AhAl3} can then be
formulated by stating that, for the considered reference frame, reduction
takes place precisely along the  $t=const$ hyperplane through $P.$

Note that this rule is manifestly covariant and it is quite natural, since
the scope of statevector reduction is to allow one to exploit the
information gained in a measurement for evaluating the probabilities of
future (for
him) events. However, taking such a position has far reaching consequences and
requires a radical revision of the meaning of the wavefunction. Namely, the
wavefunction cannot any longer be seen as a function on the space-time
continuum
but it becomes a function on the set of space-like hypersurfaces: the value of
$\Psi $ at a space-time point depends, in general, on the particular space-time
surface crossing the point we take into account.  We should then write,
in place of $\Psi (x,t),$ $\Psi =\Psi (\sigma ,x),$ where $x$ runs over the
points
of the hypersurface $\sigma .$

An intuitive understanding of this fact can be obtained by considering
(Fig.3) an objective space-time point $P$ at which a local position
measurement occurs for a particle in a state which is different from zero
along two distinct world lines ($l_{1}$ and $l_{2})$.
\begin{figure}[htb]
\begin{center}
{\includegraphics[scale=0.5]{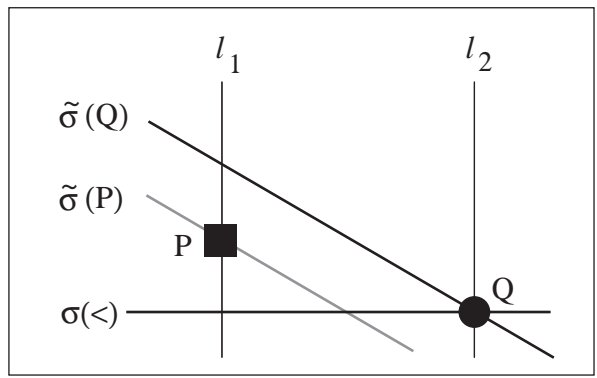}}
\caption{Illustration of the fact that in a relativistic context with
reductions the
statevector cannot be considered as a function on space time when local
measurements
are taken into account.}
\end{center}
\end{figure}
For a given reference
frame, we can consider the set of  $t=const$ hyperplanes. Then, for one of
such hyperplanes $\sigma (<)$ associated to a time prior to the one
characterizing the hyperplane through $P$, the wavefunction has a value
different from zero at the point $Q$ in which  $\sigma (<)$ intersects  $%
l_{2}:$ $\Psi ({\sigma }(<),Q) \neq {0}.$ We can now consider another reference
frame such that in it, the $ t=const$ hyperplane $\widetilde{\sigma }(Q)$
through
$Q$ is characterized by a time label greater than the one characterizing the
hyperplane $%
\widetilde{\sigma }(P)$ of the same family going through $P.$ If we suppose
that in the local measurement the particle is detected at $P$ (note that
this is
an objective, i.e., reference frame independent statement), then we have:
$\Psi(
\widetilde{\sigma }(Q),Q)=0.$ Thus, according to the Aharonov and Albert
proposal,
the wavefunction must take different values at the same space-time point
according to
which space-like surface passing through it we take into account.

We conclude this section by remarking that the idea that
to deal with relativistic quantum systems one must associate different
statevectors to different space-like hypersurfaces is a quite hold one,
going back to the fundamental papers by Tomonaga and Schwinger \cite{ToSc}.
However, within the formalism proposed by these authors, the expectation
value of any local observable having its compact support contained in the
common part of two hypersurfaces (and thus in particular the value of the
wavefunction at a point in which two such surfaces intersect) does not
depend on which of the surfaces one is taking into account. This is not
surprising: the Tomonaga-Schwinger theory is intended to describe (within a
relativistic context) the linear and deterministic evolution of statevectors
and does not pretend to account for nonlinear and stochastic processes like
those we are considering here, i.e., the reduction processes.

\section{Relativistic dynamical reduction theories}

As repeatedly stressed in the previous sections, Aharonov and Albert, in their
fundamental papers, have identified the crucial problems of relativistic
reduction
processes and have given clear hints about the features of any theory which
should account
for their occurrence. But they have not proposed any explicit and
consistent dynamical
mechanism specifying what occurs during a measurement process. Identifying
such a
new (universal)
 mechanism is precisely the  aim of dynamical reduction theories.

\subsection{The GRW theory}

The first consistent and precise proposal of a dynamical theory accounting,
at the nonrelativistic level, for the nonlinear and stochastic reduction
process is the so called GRW theory \cite{nGRW}. As pointed out by Bell \cite
{NBell}, this approach corresponds to accepting that \textit{Schr\"{o}%
dinger's equation is not always right }and taking into consideration
stochastic and nonlinear modifications of it which allow a unified treatment
of all natural processes including  the objectification of macroscopic
properties. The
model is based on the assumption that besides the standard quantum
evolution, physical systems are subjected to spontaneous localizations
occurring at random times with an appropriate average frequency and
affecting their elementary constituents. Such processes are formally
described in the following way.

Let us consider a system of $N$ particles. When the $i$-th particle of the
system suffers a localization, the wavefunction changes according to
\begin{equation}
\Psi ({\mathbf{r}}_{1},...,{\mathbf{r}}_{N})\Rightarrow \Psi _{\mathbf{x}}(%
{\mathbf{r}}_{1},...,{\mathbf{r}}_{N})=\frac{\Phi _{\mathbf{x}}({\mathbf{r}}%
_{1},...,{\mathbf{r}}_{N})}{\left\| \Phi _{\mathbf{x}}\right\| }  \label{n1}
\end{equation}
\begin{equation}
\Phi _{\mathbf{x}}({\mathbf{r}}_{1},...,{\mathbf{r}}_{N})=\left(
\frac{\alpha }{%
\pi }\right) ^{\frac{3}{4}}e^{-\frac{\alpha }{2}({\mathbf{r}}_{i}-{\mathbf{x}}%
)^{2}}\Psi ({\mathbf{r}}_{1},...,{\mathbf{r}}_{N}).  \label{n2}
\end{equation}
The probability density of the process occurring at point \textbf{x} is
given by $\left\| \Phi _{\mathbf{x}}\right\| ^{2}.$ For what concerns the
temporal aspects of the process we assume that the localizations of the
various constituents (particles) occur independently at randomly
distributed times with a mean frequency $\lambda _{m}$ which depends on their
mass. We choose $\lambda_{m} = \lambda (m/m_{0})$, where $m$ is the
mass of the particle, $m_{0}$ the nucleon mass and $\lambda $ is of the
order of $10^{-16}\sec ^{-1}.$ The localization parameter $1/{\sqrt{
\alpha }}$ is assumed to take the value $10^{-5}cm$.

As the reader can easily grasp the model does not entail any appreciable
deviation from standard quantum mechanics for microsystems since such systems
suffer, on the average, one localization every $10^{9}$ years. The most
appealing feature of the model derives from its trigger (or amplification)
mechanism: in the case of a macroscopic system the frequency of the
localizations
is amplified with the number of its constituents and, in the case of an
almost rigid body, each localization process amounts to a localization of
the centre of mass, so that superpositions corresponding to different
locations of a macroscopic object are suppressed in about $10^{-7}\sec $.

The physically relevant features of the just described model for what
concerns the reduction process (the one in which we are primarily
interested in
here) should be clear: if one wants to get information about an observable
(a property) of a microsystem, one has to use it to trigger a macroscopic
change. Different \textit{outcomes} of the \textit{measurement} are
then correlated to different positions of some macroscopic system (to be
precise,
to macroscopically different mass distributions). But the theory does not
tolerate the formation of  superpositions of such macroscopically different
states, leading, just as a consequence of the universal dynamics ruling all
natural processes, to a definite outcome. Summarizing: the interaction
between the measured system and the measuring apparatus strives to create
superpositions of macroscopically different states, but the dynamics forbids
such processes - measurements have outcomes in extremely short times.

\subsection{The CSL theory}

The model of the previous subsection, even though it contains all the
essential elements allowing to
overcome the problems affecting the quantum theory of measurement, has a
drawback: it does not
preserve the symmetry requirements of quantum mechanics for identical
constituents. One could
easily circumvent this difficulty by a theoretical scheme quite similar to
the one presented above, but a more elegant (even though physically
equivalent) formalism
(CSL) based on a stochastic evolution equation for the statevector has been
worked out
\cite{nPearle},
\cite{nGPR}. Let us list its basic features. The
evolution equation is:
\begin{equation}
\frac{d\left| \Psi _{w}(t)\right\rangle }{dt}=\left[ -\frac{i}{\hbar }%
H+\sum_{i}A_{i}w_{i}(t)-\gamma \sum_{i}A_{i}^{2}\right] \left| \Psi
_{w}(t)\right\rangle .  \label{n3}
\end{equation}
In equation (5.3) $H$ is the hamiltonian, the quantities $A_{i}$ are a set of
commuting self-adjoint operators while $w_{i}(t)$ are c-number
Gaussian stochastic processes satisfying:
\begin{equation}
\left\langle \left\langle w_{i}(t)\right\rangle \right\rangle =0,\quad
\left\langle \left\langle w_{i}(t)w_{j}(t^{\prime })\right\rangle
\right\rangle =\gamma \delta _{ij}\delta (t-t^{\prime }).  \label{n4}
\end{equation}
For the moment, let us assume that the operators $A_{i}$ have a purely
discrete spectrum and let us denote as $P_{\sigma }$ the projection
operators on their common eigenmanifolds. The precise way in which the model
works is defined by the following prescription: if a homogeneous ensemble
(pure case) is associated at the initial time $t=0$ to the statevector $%
\left| \Psi (0)\right\rangle ,$ then the ensemble at time $t$ is the union
of homogeneous ensembles associated with the normalized statevectors
$\left| \Psi _{w}(t)\right\rangle /\left\| \left| \Psi _{w}(t)\right\rangle
\right\| ,$ where $\left| \Psi _{w}(t)\right\rangle $ is the solution of
the evolution equation with the assigned initial condition and for the
specific stochastic process $w$ which has occurred in the interval $(0,t).$
The probability density for such a subensemble is the {\it cooked} one,
given by:
\begin{equation}
P_{cooked}[w]=P_{Raw}[w]\left\| \left| \Psi _{w}(t)\right\rangle \right\|
^{2},  \label{n5}
\end{equation}
where
\begin{equation}
P_{raw}[w]=\frac{1}{N}e^{-\frac{1}{2\gamma }\sum_{i}\int_{0}^{t}d\tau
w_{i}^{2}(\tau )},  \label{n6}
\end{equation}
$N$ being a normalization factor. It is easy to show that the just described
dynamical process, when the non-hamiltonian part dominates
the hamiltonian one, drives each statevector within one of the eigenmanifolds
associated to the projection operators  $P_{\sigma },$ with the appropriate
probabilities.

To get a theory which leads to the desired reductions (i.e. which works like
the GRW theory) one has to make a choice for the operators $A_{i}$ which is
directly suggested by the GRW theory itself. It is obtained by identifying the
discrete index $i$ with the continuos index \textbf{r} and the above
operators with an appropriately averaged mass density operator $M(\mathbf{r}%
):$%
\begin{equation}
M({\mathbf{r}})=\sum_{k}m^{(k)}N^{(k)}(\mathbf{r}),  \label{n7}
\end{equation}
\begin{equation}
N^{(k)}({\mathbf{r}})=\left[ \frac{\alpha }{2\pi }\right]
^{\frac{3}{2}}\sum_{s}\int
d{\mathbf{q}}e^{-\frac{\alpha
}{2}({\mathbf{q}}-{\mathbf{r}})^{2}}a_{(k)}^{\dagger
}({\mathbf{q}},s)a_{(k)}({\mathbf{q}},s),  \label{n8}
\end{equation}
where  $a_{(k)}^{\dagger }({\mathbf{q}},s)$ and $a_{(k)}(%
{\mathbf{q}},s)$ are the creation and annihilation operators of a particle of
type $k$ ($k=$ proton, neutron, electron,...) at a point \textbf{q}, with
spin component $s.$ The equations replacing (5.3) and (5.4) are then:
\begin{equation}
\frac{d|\Psi _{w}(t)>}{dt}=\left[ -\frac{i}{\hbar }H+\int d{\mathbf{r}}M(%
{\mathbf{r}})\widetilde{w}({\mathbf{r}},t)-\frac{\gamma }{m_{0}^{2}}\int d%
{\mathbf{r}}M^{2}({\mathbf{r}})\right] |\Psi _{w}(t)>  \label{k1}
\end{equation}
and
\begin{equation}
\left\langle \left\langle \widetilde{w}({\mathbf{r}},t)\right\rangle
\right\rangle =0,\quad \left\langle \left\langle \widetilde{w}({\mathbf{r}},t)%
\widetilde{w}({\mathbf{r}}^{\prime },t^{\prime })\right\rangle \right\rangle =%
\frac{\gamma }{m_{0}^{2}}\delta ({\mathbf{r}}-{\mathbf{r}}^{\prime })\delta
(t-t^{\prime }).  \label{k2}
\end{equation}

With these choices, when the parameter
$\gamma
$ is related to those of the GRW theory according to  $\gamma =\lambda \left(
{4\pi }/{\alpha }\right) ^{\frac{3}{2}}$,  the
theory implies that any macroscopic object is always extremely well
localized in space and that if any interaction leads, as a consequence
of the hamiltonian dynamics, to a superposition of differently located
states of a
macroscopic system, then reduction takes place almost immediately to one of
them, with the
appropriate probability.

\subsection{Relativistic CSL}

The first attempt to get a relativistic generalization of the CSL theory has
been performed by P. Pearle\cite{P}. A detailed investigation of the
model proposed by him together with a discussion of all its most relevant
features
has been presented in ref.\cite{GGP1}. In particular, in this last paper the
relativistic stochastic invariance of the model, as well as the
integrability of its evolution equation have been analyzed in all details.
For the moment, let us simply recall the general scheme. One adopts a
Tomonaga-Schwinger approach within a quantum field theoretical framework and
one accounts for the stochasticity of the evolution by an appropriate
stochastic
interaction term. Accordingly, one considers a Lagrangian density
\begin{equation}
L(x)=L_{0}(x)+L_{I}(x)V(x)  \label{n9}
\end{equation}
where $L_{0}(x)$ and $L_{I}(x)$ are scalar functions of the fields, $L_{I}(x)
$ does not contain derivative couplings, and $V(x)$ is a \textit{c}-number
stochastic process which is a scalar for Poincar\'{e} transformations. One
chooses for $V(x)$ a Gaussian noise with mean zero. In order to have
relativistic stochastic invariance, its covariance must be an invariant
function:
\begin{equation}
\left\langle \left\langle V(x)V(x^{\prime })\right\rangle \right\rangle
=A(x-x^{\prime }),\qquad A(\Lambda ^{-1}x)=A(x).  \label{n10}
\end{equation}
In refs.\cite{P} and \cite{GGP1} the choice
\begin{equation}
A(x)=\lambda \delta (x)  \label{n11}
\end{equation}
has been made. It has to be stressed that such a choice, due to its white
nature in time, gives rise to specific problems related to the appearence of
untractable divergences. On the other hand, if one would not
require the function $A(x)$ to be white in time various unacceptable
consequences would
emerge \cite{GGP1}. Thus, we plainly accept that the program of a relativistic
generalization of dynamical reduction theories is still an open one.
However, the just
mentioned problems are of technical nature, and one can hope to succeed in
overcoming
them. On the other hand, as we will show, the formal structure of the
theory is perfectly
satisfactory. Accordingly, we will disregard here this kind of difficulties
and we will
concentrate our attention on the general features of the theory.

We still have to define the precise dynamics of the model. This can be
summarized in the following terms:

\begin{itemize}
\item   The fields are solutions of the Heisemberg equations associated to $%
L_{0}(x).$

\item   The statevector obeys the evolution equation
\begin{equation}
\frac{\delta |\Psi _{V}(\sigma )>}{\delta \sigma (x)}=\left[
L_{I}(x)V(x)-L_{I}^{2}(x)\right] |\Psi _{V}(\sigma )>.  \label{n12}
\end{equation}
\end{itemize}

Note the skew-hermitian character of the coupling to the stochastic field.
This equation, just as those of nonrelativistic CSL, does not preserve the
norm of the statevector but it preserves its average value. Accordingly, one
has to introduce an appropriate ``cooking'' procedure \cite{GGP1} for the
probability
of occurrence of a given potential which parallels the one (\ref{n5}) of CSL.

In the specific case of refs.\cite{P} and \cite{GGP1} consideration has
been given to a hermitian scalar meson field $\Phi (x)$ coupled to a fermion
field $\Psi (x)$, and the following choices have been made for the Lagrangian
densities:
\begin{equation}
L_{0}(x)=\frac{1}{2}\left[ \partial _{\mu }\Phi (x)\partial ^{\mu }\Phi
(x)-m^{2}\Phi ^{2}(x)\right] +\overline{\Psi }(x)\left[ i\gamma ^{\mu }%
\partial _{\mu }-M\right] \Psi (x)+\eta \overline{\Psi }(x)\Psi (x)\Phi (x)
\label{n13}
\end{equation}
\begin{equation}
L_{I}(x)=\Phi (x).  \label{n14}
\end{equation}

The physically relevant features of the model can be easily understood. The
nonlinear and stochastic nature of eq.(\ref{n12}) implies that the dynamics
leads
to the suppression of superpositions of different states of the meson
field. This,
if one takes into account that fermions which are differently located are
associated with different mesonic clouds, makes clear that the dynamics leads
(indirectly) to a (very unfrequent) localization of the fermions. For lack
of space we
cannot analyze the theory in all its details and
we refer the reader to ref.\cite{GGP1} for an exhaustive discussion of all
its features.

What matters for our analysis is just the characteristic of the theory of
leading to the localization of fermions, in particular of  nucleons.
Actually, by appropriately choosing the constants of the model, one shows
(ignoring the problem of the divergences) that in the nonrelativistic
limit the theory exhibits features which are extremely similar to those of
the CSL model.

Two remarks seem appropriate:

\begin{itemize}
\item   Due to the nonhermitian structure of the evolution operator, the
states associated to different hypersurfaces can take different values at a
given objective space-time point (equivalently, the theory attaches different
expectation values to operators having a common compact support). This fact
parallels strictly the features which have been identified by Aharonov and
Albert \cite{AhAl2} and which have been discussed in section 4.

\item  The theory, just as the GRW and CSL models, does not exhibit
stochastic time reversal invariance: only the forward time
translation  Poincar\'{e} semigroup is represented. Accordingly,
the initial conditions must be given on a precise, objective space-like
hypersurface. The transformation from one observer to another must be
discussed by adopting the passive point of view.
\end{itemize}

We conclude this subsection by discussing an explicit example of a
measurement--like process. To this purpose it is sufficient  to take
into account once more a microscopic system possessing an internal degree
of freedom
which will be treated quantum mechanically, while one  treats  its space
degrees of freedom as  classical (in particular one  disregards
the spreading of wavepackets and one  speaks of the world lines of the system).
One also assumes that  a macroscopic system enters into the game and that it
also can be treated, for what concerns its spatial degrees of freedom, in
classical terms. The world lines of the micro and the macroscopic systems are
assumed to intersect at a precise objective space-time point. The
macroscopic system mimics an apparatus measuring an observable of the
internal Hilbert space of the microscopic system. The system-apparatus
interaction is assumed to induce, according to the internal microstates of
the system, different displacements of a macroscopic part (the pointer) of
the apparatus. Such a mobile part of the device contains, obviously, an
enormous number of nucleons (of the order of Avogadro's number). They would
end up in a superposition of states corresponding to different positions if
no reducing
dynamics would be effective. But, as we have made plausible, different
locations of a
nucleon are (very seldom) suppressed by the nonlinear evolution which does
not tolerate
superpositions of the associated different mesonic clouds. An amplification
mechanism mirroring precisely the one of the GRW and CSL theories is present
also in the relativistic model we are discussing. This means that the
relativistic dynamics leads to the suppression of all but one the
possible final macroscopically different configurations in an extremely
short interval of
the proper time of the apparatus.

We have described the process in physical terms. Let us now look at it in
mathematical terms and in a language which does not mention
different reference frames but only objective points or surfaces of the
space-time continuum. We have a space-like surface $\sigma _{0}$ (to be
identified with the one chosen for accounting of the big bang?) on which the
initial conditions for the system and the apparatus are given. The
dynamical equation
determines in a unique way the statevector on any space-like surface lying
entirely in the
future of  $\sigma _{0}.$ The previous discussion should have made clear
that if we change the space-like surface we are interested in (e.g. by
considering smooth continuous variations of it), it is just when we cross
the point in which the system-apparatus interaction takes place that the
nonlinear dynamics becomes effective and leads to a macroscopic definite
state of the pointer (and, obviously, to a corresponding precise microstate
of the system). Note that what matters for the reduction process
 is the fact that the considered space-like surface crosses an
objective space-time point and not the precise way in which it is changed.
If we cross the
point by modifying the whole space-like surface we are considering (for
instance by
translating it rigidly in the direction of the time axis) or if we keep a
far-away part of
the surface fixed and we modify it only around the space-time point at
which  the
system-apparatus interaction occurs, the change of the statevector is
(objectively) the same\footnote{%
Obviously, the same objective situation of the pointer as well as the
observable which becomes definite for the system will be described in
different terms by different observers. For instance if one observer adopts
a certain reference frame he could claim that the pointer is aligned with his
z-axis, while a rotated observer will claim that it is aligned with his
x-axis. But this is totally irrelevant: we are not making reference to the
language used by the observers, to the labels they attach to different
space-time points and so on, we are specifying what happens in a covariant
language, such as the one making reference to the objective point in which
the system apparatus interaction takes place or to the fact that such a
point belongs or does not belong to the volume between the initial
space-like surface and the one we are interested in.}: the pointer, in an
extremely short time, acquires a precise \textit{final} position differing
from the initial \textit{ready to measure} one, a position which is
correlated to a precise final state of the measured system, i.e. the one
corresponding to the \textit{measurement outcome}.

At this point the reader should have perfectly clear that the basic features
of the model under discussion are precisely those which have been identified
in refs.\cite{AhAl2} and \cite{AhAl3}, in particular that different
statevectors are attached to different hypersurfaces, that the \textit{value}
of the wavefunction can differ at the same space-time point according to the
space-like hypersurface going through that point one takes into account, and
that, in a quite precise sense, reduction occurs, for all observers,
along the hypersurfaces going through the point at which the
measurement process takes place.

In what follows we will reconsider the reduction process induced by the
theory with
reference to an oversimplified model of two correlated microsystems
subjected to
measurements in space-like separated regions, an analysis which will allow
us to clarify some relevant features connected with quantum nonlocality and
with the use of counterfactual arguments within a relativistic and nonlocal
context. Now it is time to devote particular attention to the problem of the
properties objectively possessed by individual physical systems within a
theoretical framework like the one we have described in this section.

\section{Properties and events in the relativistic context}

We are ready to tackle a problem of great conceptual relevance, i.e., the
one of the emergence of definite properties of an individual physical system
from the sea of the\textit{\ potentialities} which characterize it within
the quantum framework. Obviously, this problem has extremely close
connections with the \textit{measurement} or the\textit{\
macro-objectification} problem. For these reasons it is obvious that taking
a precise position about the reduction process, such as the one entailed by
relativistic CSL, puts precise limitations concerning the
situations in which one can consistently speak of properties objectively
possessed by an individual physical system at a give space-time point of its
world line.

Before proceeding it is useful to reconsider shortly the same problem within
nonrelativistic quantum mechanics. In such a context there is an absolute
time, so that one can discuss the question of the properties objectively
possessed by a physical system  at a given instant $t.$
Suppose we have a system $S$ (elementary or composed of various
constituents), whose state is represented, at the considered time, by the
statevector $|\Psi (t)>.$ The standard wisdom about properties (or in
Einstein's language about the\textit{\ elements of physical reality}) leads
to a quite natural assumption:

\begin{quotation}
If consideration is given to an observable whose associated self-adjoint
operator is $\Omega ,$ we claim that the events ``the considered observable
is definite"
and  ``the system
$S$ possesses the objective property $\Omega =\omega _{k}$" occur \textit{iff}
$|\Psi (t)>$ is an eigenstate of $\Omega $ belonging to the indicated
eigenvalue. If
this is not the case, we claim that the event ``the considered observable is
indefinite" occurs.
\end{quotation}

Some remarks are appropriate.

\begin{itemize}
\item  If one accepts, as it is usual within
textbook quantum mechanics, that all (bounded) self-adjoint operators
correspond to physically measurable quantities, then any system S,
considered as a whole, possesses always some properties. In fact in
any case there is at least one self-adjoint operator such that $|\Psi (t)>$
is one of
its eigenstates (the most trivial example being the projection operator $%
|\Psi (t)><\Psi (t)|$ itself).

\item  The above position concerning the possibility of speaking of objective
properties matches the well known fact that the theory allows one to make
the counterfactual statement: if at time $t$ a measurement process of $%
\Omega $ were performed it would yield with certainity the outcome $%
\omega _{k}.$ For an analysis of this point, see the remarks below.

\item  If the assumption under the first item above is satisfied, then there
are certainly physical observables which do not have a definite value, in
the sense that the theory attaches genuinely nonepistemic nonvanishing
probabilities to different outcomes in a prospective measurement of such
observables at
the considered time.
\end{itemize}

To analyze in greater details the just discussed situation let us consider
the case of one particle which is in the (improper) superposition of two
position eigenstates \footnote{Obviously, to be correct, one should
consider in place of
the states $|x_{1}>$ and  $|x_{2}>$ normalized wavefunctions different from
zero only in
extremely small intervals around the indicated positions and
correspondingly a detector
whose acceptance window is larger than the extension of the wavepacket
around $x_{2}$.}:
\begin{equation}
|\Psi >=\frac{1}{\sqrt{2}}\left[ |x_{1}>+|x_{2}>\right] .  \label{a1}
\end{equation}
For such a state there is no matter of fact about the location
of the particle: its position is indefinite. Suppose now that at time $t$ a
detector is
placed at point
$x_{2}$ and it does not detect the particle. Immediately after $t$ the
state, in
accordance with the reduction postulate, becomes $|x_{1}>$ so that we
can legitimately claim that the definite objective property of the particle
being
at $x_{1}$ has emerged as a consequence of the measurement at $x_{2}.$ Note
that this
statement is perfectly legitimate both from the point of view of the above
criterion
concerning possessed properties as well as from the one of counterfactual
reasoning. In
fact, the formalization of such reasonings requires to define the
accessibility sphere from
the actual world, and it is common practice in  physics to consider as
accessible (i.e. as those nearest to the actual one) those worlds in which the
physical laws are the same as those of the actual world, and which
coincide, up to the
time one is interested in, with the actual world itself. Accordingly, in
our case the
accessibility sphere is represented by those worlds in which quantum
mechanics holds
and for which the state of the physical system we are interested in
coincides up to a
time following immediately  $t$ with the one describing it in the actual
world. In
particular in all accessible worlds the premises: the particle has not been
detected
at $x_{2}$ and, accordingly, its state is $|x_{1}>$, are true.

As the reader should have clear, the situation changes radically in a
relativistic
context. In fact, as repeatedly stressed, the value of the wavefunction at
the space-time
point $(x_{1},t)$ (i.e. in our case the fact that it differs from zero
either only
around  point
$x_{1}$ or around both points) depends not only on the considered
(objective) space-time point,
but also on the space-like surface through  it which we take into account.
How should one proceed
to make statements about the property related to the location of the
particle at a given
time? The first important thing which has to be remarked is that in the
spirit of an
approach like the one of CSL bearing on {\it reality} as opposed to {\it
intersubjective
appearences}, the prescription which should lead to the legitimate
conclusion that a property
is definite and possessed or it is indefinite, must be covariant, i.e.,
independent from the
reference frame one is considering. If one takes into account that the
theory attaches a
precise statevector to any space-like surface and that one is interested in
a statement
concerning an objective space-time point $P$ on the world line of the
system we are
considering, there are only two natural covariant prescriptions satisfying
the above
conditions, and they make reference to the past or to the future light cone
from
$P$, respectively. For reasons which are easily understandable (more about
this in what
follows) it is appropriate to resort to the criterion which takes into
account the past light
cone. Thus we replace the previous assumption about the attribution of
objective
properties to an individual physical system (at a point $P$ of its world
line) by the following one:

\begin{quotation}
If consideration is given to a space-time point $P$ and to a given
observable whose
associated self-adjoint operator is $\Omega ,$ we consider, first of all,
the space-like
surface $\sigma (P)$, constituted by the past light cone from
$P$ and the part outside it of the initial surface $\sigma_{0}$. The theory
assignes a
precise statevector $|\Psi(\sigma (P))>$ to such a surface. Then we proceed
as before,
i.e., we claim that the system
$S$ possesses the objective property ``$\Omega =\omega _{k}"$ \textit{iff}
 $|\Psi(\sigma(P))>$ is an eigenstate of $\Omega $ belonging to the
indicated eigenvalue, and,
if this is not the case, we claim that the considered observable is indefinite.
\end{quotation}

Once more some remarks are appropriate:

\begin{itemize}
\item The above criterion implies that the statement
``the particle has the definite property $\Omega =\omega _{k}$" is correct
when, in the
past light cone from $P$ an appropriate preparation procedure or an interaction
corresponding to the measurement of
$\Omega$ has occurred. Note that for an elementary particle this means that
there is a
point along its world line in which it has interacted with an appropriate
device,
while in more general cases like the one of a pair of far-away correlated
particles in
an entangled state, the property of one particle becomes definite also when
in its past
light cone there is a point in which a measurement of the relevant
observable  has been
performed on the correlated particle.
\item The above position concerning the possibility of speaking of objective
properties matches the  fact that the theory allows one to make
the counterfactual statement ``if at the space-time point $P$  a
measurement process of the
observable $\Omega$ were performed it would yield the outcome
$\Omega_{k}$", provided
one makes the perfectly reasonable and covariant assumption that the
accessible worlds
from the actual one are those in which the laws of nature are the same as
those of the
actual world and the physical situation matches exactly the actual one
within the past
light cone from $P$.
\end{itemize}

\section{Relativistic reduction and nonlocality}

The fundamental issues discussed in the previous section become
particularly interesting in
connection with quantum nonlocality. To investigate this subtle point it is
particularly
useful to resort to an oversimplified model which exhibits all relevant
features of
relativistic CSL (in particular it is relativistically invariant,
stochastic, nonlinear and
nonlocal) but it is remarkably more simple. Such a model has been already
considered in
the previous sections, but, before proceeding, it is appropriate to make it
absolutely
precise.

\subsection{Preliminary considerations: a relativistic toy model}

We will deal with one or two particles, each having (as in the previous
sections) space degrees of freedom obeying a classical relativistic
dynamics, plus a quantum internal degree of freedom which behaves like a
scalar under Lorentz transformations and which we will identify, for
simplicity, with the isospin space of an isospin $1/2$ particle. We will
consider
an operator of this space with eigenvalues +1 and --1. When we will deal with
two particles we will correspondingly consider two such operators ,
$\Theta ^{(i)}$ (i=1,2) and we will denote by $|i,+>,$ $|i,->$ the
corresponding eigenvectors. Obviously, in the internal space of a
particle one can consider any two by two hermitian matrix and in the case
of two particles the full algebra of hermitian operators in the four
dimensional internal space. However, for our purposes (as in the modern
versions of the EPR argument) there will never be the need to resort to
noncommuting observables referring to a particle, so that we will always deal
with the operators $\Theta ^{(i)}$ .

In the theory, besides the two particles (1,2), there are objects (\textit{A},%
\textit{B}, ... ), simulating apparata measuring the observables $\Theta
^{(i)}$, which are characterized by space degrees of freedom obeying a
classical relativistic dynamics and by three possible internal states, $%
(r,+,-)$. Since they will always be in one of these three states it is not
relevant, for the present analysis, to be precise about the nature of their
internal space. In particular, one could consider the states $|r>$, $|+>$
and $|->$ as three ortogonal vectors in a three-dimensional Hilbert space or
as three classical \textit{labels}, which are Lorentz scalars. The objects,
even though representing measuring devices, are supposed to be point-like,
so that we can consider their world lines (which we will not draw in the
figures) and the space-time points (which will be represented by small black
squares) in which such world lines intersect the particles' world lines.
Moreover the objects (\textit{A},\textit{B}, ... ) are characterized by
parameters $g_{A},g_{B},$ etc., which can take one of two values \{0,1\}
(chosen at free-will by an experimenter) corresponding to the apparatus
being ``switched on'' or ``switched off'' respectively.

\subsection{The one-particle case}

To warm up we begin by discussing the case of one particle. We consider its
world line originating from the space-like surface $\sigma _{0}$ on which
the initial conditions are given and, with reference to the internal degree
of freedom, we assign the statevector on this surface by expressing it as a
linear superposition of the eigenstates of the operator $\Theta $ according
to:
\begin{equation}
|\Psi (\sigma _{0})>=\alpha |+>+\beta |->.  \label{35}
\end{equation}

- \textit{The completeness assumption} is embodied in the assertion that the
assignement of the initial state (\ref{35}) (besides the relativistic
classical dynamics for the propagation of the free particle) represents the
maximum of information we can have about the particle itself and determines
all what we can know about it.

- \textit{The experimental context}: along the world line of the particle,
at the space-time point $R$, there is an apparatus $A$ devised to measure $%
\Theta $, which can be switched off or on.

- \textit{Dynamics}: it is nonlinear and stochastic. The theory associates
to any space-like surface $\sigma $ in the future of $\sigma _{0}$ a
statevector according to the following rules\footnote{%
As already remarked, we will never consider other observables besides $%
\Theta $. However, an exhaustive theory should deal with all hermitian
operators in the internal space. The reader will have no difficulty in
generalizing the rules to cover such a case. The whole procedure requires
only to express the initial state as a linear combination of the eigenstates
of the observables one is interested in.}:

Denote by $V(\sigma ,\sigma _{0})$ the space-time volume enclosed by the two
indicated surfaces.

a). If
\begin{equation}
\left[ R\notin V(\sigma ,\sigma _{0})\right] \vee \left[ g_{A}=0\right] ,
\label{36}
\end{equation}
then the state $|\Psi (\sigma )>$ associated to the surface is
\begin{equation}
|\Psi (\sigma )>=|\Psi (\sigma _{0})>,  \label{37}
\end{equation}
while

b). if
\begin{equation}
\left[ R\in V(\sigma ,\sigma _{0})\right] \wedge \left[ g_{A}=1\right] ,
\label{38}
\end{equation}
the state is
\begin{equation}
either\quad |\Psi (\sigma )>=|+>\quad or\quad |\Psi (\sigma )>=|->,
\label{39}
\end{equation}
the two alternatives occurring randomly with probabilities $\left| \alpha
\right| ^{2}$ and $\left| \beta \right| ^{2}$ , respectively. Thus, when a
spacelike surface crosses the region in which an apparatus is switched on a
real dice-playing leading to one among two possible states takes place: the
probabilities governing the process have a nonepistemic status.

In equations (\ref{35}), (\ref{37}) and (\ref{39}) we have skipped the
indication of the apparatus state but it is understood that the apparatus
will be in one of the states (r,+,-) matching the one of the system.

A final comment is appropriate. The dynamics satisfies the consistency
requirement that considering the evolution leading from $\sigma _{0}$ to $%
\sigma $ and then the one leading from $\sigma $ to $\sigma _{1}$ is the
same as going directly from $\sigma _{0}$ to $\sigma _{1}.$ Thus, if one
considers the case b) and assumes that along the particle's world line
there is another apparatus which is switched on and devised to measure the
same observable at a point $\widetilde{R}$ in the future of $R$, then the
statevector to be assigned to surfaces $\sigma $ such that $\widetilde{R}\in
V(\sigma ,\sigma _{0})$ would be the same as the one appearing in (\ref{39}).

\subsection{Events in the toy model: the one-particle case}

In the one-particle case of subsection (7.2) let us  consider  a
point  \textit{P} preceeding, on the world line of the particle, the point
\textit{R}. In such a case, since the precise rules of the model tell us
that the state on the space-like surface $\sigma (P)$ is not an eigenstate
of $\Theta $ , the specific event ``the associated property is
indefinite'' occurs.

Obviously, if consideration is given to a point $\widetilde{P}$ following%
\textit{R} when $g_{A}=1$, then the surface $\sigma (\widetilde{P})$ is
such that for it condition (\ref{38}) holds. Accordingly, as shown by Eqs.(%
\ref{39}), the statevector is an eigenstate of $\Theta $ (which one between $%
|+>$ and $|->$ is determined by the genuinely stochastic process taking
place at $R$) and the the specific event ``the property associated
to $\Theta $ is definite and equals + (-)'' occurs.

The case under discussion, since it does not involve space-like separated
events and nonlocal effects, does not raise any specific problem differing
significantly from those which characterize also nonrelativistic quantum
mechanics, i.e. the fact that one cannot avoid to take into account the
event ``there is no property referring to $\Theta $ ''. The situation
changes remarkably in the two particle case we are going to discuss.

\subsection{The two-particle case}

We start by generalizing the toy dynamics to the case in which there are two
particles in place of one. First of all we have to specify the world lines
describing the classical motion of the two particles and we must assign the
statevector referring to the internal degrees of freedom on the initial
space-like surface. To be general we should express it as an arbitrary
normalized linear combination of the four ortonormal states $|i,s>\otimes
|j,t>$ ($i,j=1,2;s,t=+,-$) which are the common eigenvectors of the two
commuting operators $\Theta ^{(k)}$. Similarly we should take into account
the possibility of measuring, for each particle, operators which do not
commute with those considered above. Moreover also the case of correlation
measurements involving different operators for the two particles should be
taken into account. However, for our purposes we can limit our
considerations to a very specific initial state, i.e. to the state

\begin{equation}
|\Psi (\sigma _{0})>=\frac{1}{\sqrt{2}}\left[ |1+,2->-|1-,2+>\right]
\end{equation}
and to the operators $\Theta ^{(i)}$.
\begin{figure}[htb]
\begin{center}
{\includegraphics[scale=0.75]{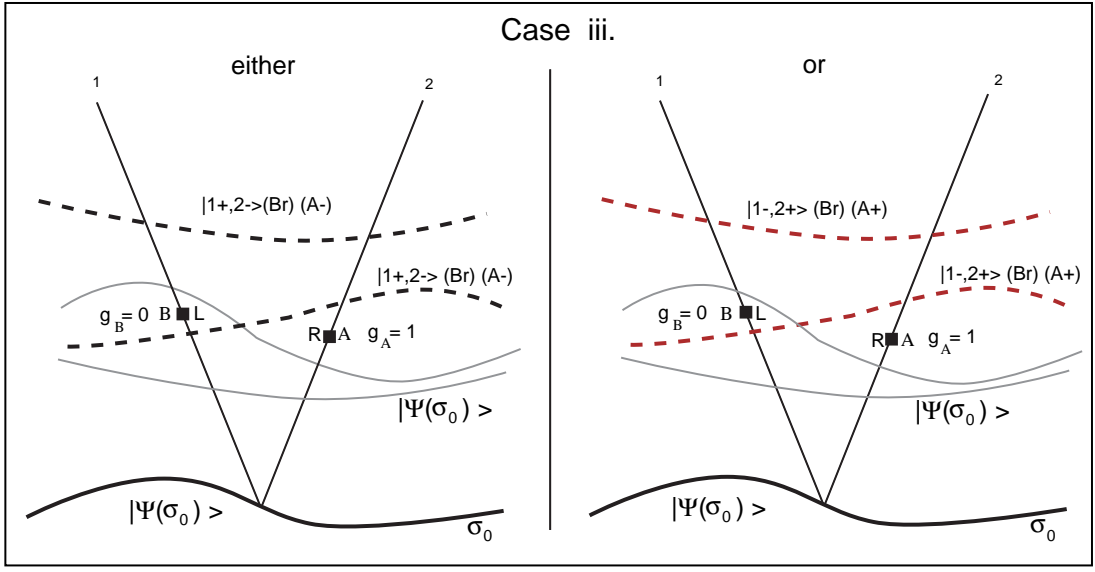}}
\caption{The dynamics of the system of two entangled particles in the
relativistic toy
model of section. 7 when only one apparatus is switched on.}
\end{center}
\end{figure}
The reader could easily generalize
our rules to arbitrary initial states and arbitrary measurement processes.
Once more we specify the rules of the game (see Figs. 4 and 5):

- \textit{Completeness}: the assignement of the initial state (7.6)
represents the maximum information one can have about the system.

- \textit{The experimental context}: along the world lines of the particles,
at two space-time points \textit{R} (at right for particle 2) and \textit{L }%
(at left for particle 1) there are two apparatuses \textit{A} and \textit{B}
(each characterized by three possible internal states r,+,-) devised to
measure $\Theta ^{(2)}$ and $\Theta ^{(1)},$ respectively . Each apparatus
can be switched on or off at the experimenter's free will.

- \textit{Dynamics}: once more it is nonlinear and stochastic and associates
to any space-like surface in the future of $\sigma _{0}$ a precise
statevector according to rules which are the natural generalization of those
of Section (7.2) (as before we denote by $V(\sigma ,\sigma _{0})$ the
space-time volume enclosed by the two indicated surfaces):

i. If
\begin{equation}
\left\{ \left[ R\notin V(\sigma ,\sigma _{0})\right] \vee \left[
g_{A}=0\right] \right\} \wedge \left\{ \left[ L\notin V(\sigma ,\sigma
_{0})\right] \vee \left[ g_{B}=0\right] \right\} ,  \label{40}
\end{equation}
then the state ${\vert}\Psi (\sigma )>$ associated to the surface
$\sigma $ is
\begin{equation}
|\Psi (\sigma )>=|\Psi (\sigma _{0})>.  \label{41}
\end{equation}
This situation occurs for the space-like surfaces represented by continuous
lines in Fig.5.

ii. If
\begin{equation}
\left\{ \left[ R\notin V(\sigma ,\sigma _{0})\right] \vee \left[
g_{A}=0\right] \right\} \wedge \left\{ \left[ L\in V(\sigma ,\sigma
_{0})\right] \wedge \left[ g_{B}=1\right] \right\} ,  \label{42}
\end{equation}
then the state is
\begin{equation}
either\qquad |\Psi (\sigma )>=|1+,2->\qquad or\qquad |\Psi (\sigma
)>=|1-,2+>,  \label{43}
\end{equation}
the two alternatives occurring at random with equal probabilities.

iii. If
\begin{equation}
\left\{ \left[ R\in V(\sigma ,\sigma _{0})\right] \wedge \left[
g_{A}=1\right] \right\} \wedge \left\{ \left[ L\notin V(\sigma ,\sigma
_{0})\right] \vee \left[ g_{B}=0\right] \right\} ,
\end{equation}
then the state is
\begin{equation}
either\qquad |\Psi (\sigma )>=|1+,2->\qquad or\qquad |\Psi (\sigma
)>=|1-,2+>,  \label{45}
\end{equation}
the two alternatives occurring at random with equal probabilities. This
situation occurs for the space-like surfaces represented by dashed lines in
Fig.4.

iv. Finally if
\begin{equation}
\left\{ \left[ R\in V(\sigma ,\sigma _{0})\right] \wedge \left[
g_{A}=1\right] \right\} \wedge \left\{ \left[ L\in V(\sigma ,\sigma
_{0})\right] \wedge \left[ g_{B}=1\right] \right\} ,  \label{47}
\end{equation}
then the state is
\begin{equation}
either\qquad |\Psi (\sigma )>=|1+,2->\qquad or\qquad |\Psi (\sigma
)>=|1-,2+>,  \label{48}
\end{equation}
the two alternatives occurring at random with equal probabilities. This case
is represented by the gray dashed lines of Fig.5.
\begin{figure}[htb]
\begin{center}
{\includegraphics[scale=0.75]{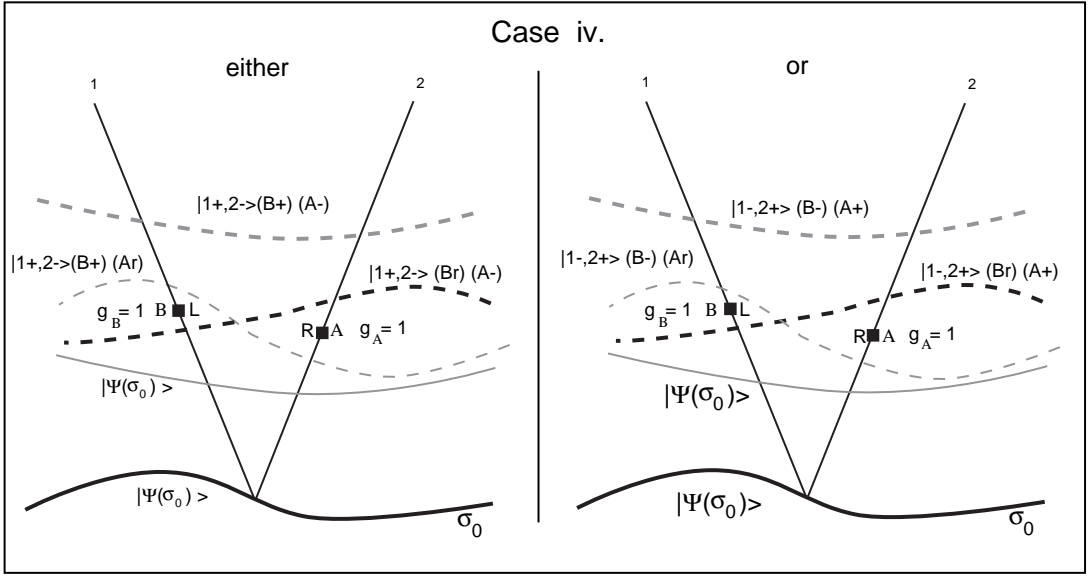}}
\caption{The dynamics of the system of two entangled particles in the
relativistic toy
model of section. 7 when both apparatuses are switched on.}
\end{center}
\end{figure}

We have depicted in Fig. 4 the two alternatives corresponding to case iii,
and in Fig. 5  those corresponding to case iv. It has to be stressed that the
two occurrences in cases ii. and iii. (when only one apparatus is on) have
no relations with the corresponding ones of case iv. In fact, leaving aside
the case in which no measurement occurs, it has to be stressed that since in
the actual world either one or both apparatuses are switched on, and the
outcomes are
genuinely stochastic, there is no definite relation between the two cases
of Fig.4
and of Fig.5 . On the other hand, in case iv., if one considers a space-like
surface, like the black dashed lines of Fig.5, passing below one of the two
points
where there is an apparatus and above the other one, and one supposes that
one of
the two alternatives has occurred, then the subsequent evolution must be
consistent with the chosen alternative, i.e., for all space-like surfaces in
the future of both \textit{A }and \textit{B}, the statevector of the system
remains the same and the previously untriggered apparatus simply registers
the property possessed by the microsystem. In this way the necessary
requirement that in any case one can consistently describe the evolution
from $\sigma _{0}$ to $\sigma _{1}$ and then the one from $\sigma _{1}$ to $%
\sigma _{2}$ is satisfied.

\subsection{Some features of the two-particle model}

The model we have just introduced has many interesting features. It contains
precise dynamical rules for assigning to each space-like surface in the
future of the surface defining the initial conditions a definite
statevector. The model is fundamentally stochastic so that, when various
alternatives can occur, they occur genuinely at random but in accordance
with precise probabilistic laws. The macroscopic apparata {\it are always in
definite states} (i.e. they always possess definite macroscopic properties)
and, for
those apparata which are switched on and for space-time points following
(on their
world lines) the event ''the microsystem triggers the apparatus'', they
match the eigenvalues of the observables they are devised to measure. In all
other instances, they correspond to their initial untriggered states.

The physical implications of the model obviously agree with those of SQM. In
fact, since taken any objective space-time point (after the system-apparatus
interaction) on the world line of an apparatus\footnote{%
These world lines are not shown in the figures, but they can be simply
thought as vertical lines in the reference frame in which the figures are
drawn, corresponding to the fact that they are at rest in this frame.} the
apparatus state is precisely defined, we can make reference to these states
to ``read'' the outcomes of the process.

Concerning its formal structure it has to be stressed that the model is
entirely formulated in a coordinate-free language and thus it satisfies the
relativistic requirements \cite{GGP1} of a stochastically Lorentz
invariant theory. In fact the statement that an objective space-time point
(the one in which there is an apparatus which is on) belongs or does not
belong to a precisely defined space-time volume is frame independent and
the internal degrees of freedom are assumed to be Lorentz scalars. If we
consider the correlations between outcomes, we see that when both
apparatuses are switched on they register either (A+) (B-) or (A-) (B+) with
equal probabilities and they never register the same outcome. Consequently
the model reproduces the perfect correlations of SQM for isospin measurements
along the same direction in the isospin singlet state.

The model satisfies the completeness requirement by assumption: there is no
better specification of the initial state than the one given by $|\Psi
(\sigma _{0})>,$ and its knowledge specifies everything about the future of
the system exception made for the actual outcomes of processes whose
probability of occurence is fundamentally nonepistemic.

Due to the fact that the model guarantees the perfect correlations of
outcomes at the two wings of the apparatus it violates Bell's locality
requirement. It is quite important to stress that:

a). The model exhibits {\it Parameter Independence}. In fact, denoting by $
P_{S}(+1|g_{R}=\alpha )$ the probability that the outcome at $S$ (taking the
values $L$,$R$) be $i$ (taking the values +1,--1) when the apparatus at $S*$
(taking the value $R$ when $S=L$ and $L$ when $S=R$) is switched off ($%
g_{S*}=0$ ) or it is on ($g_{S*}=1$ ), we have:
\begin{eqnarray}
P_{L}(+1|g_{R} &=&1)=P_{L}(+1|g_{R}=0)=\frac{1}{2} \\
P_{L}(-1|g_{R} &=&1)=P_{L}(-1|g_{R}=0)=\frac{1}{2}  \nonumber
\end{eqnarray}
and, analogously:
\begin{eqnarray}
P_{R}(+1|g_{R} &=&1)=P_{R}(+1|g_{R}=0)=\frac{1}{2}  \label{49} \\
P_{R}(-1|g_{R} &=&1)=P_{R}(-1|g_{R}=0)=\frac{1}{2}  \nonumber
\end{eqnarray}

b). The model violates {\it Outcome Independence} since the outcomes are
perfectly
correlated in spite of the fact that they have
probability 1/2 of being +1 or -1.

\subsection{Events in the two-particle case}

At this point the reader should already have perfectly clear all the
implications of the model. Suppose one is interested in an event concerning
a space-time point $P$ of the world line of the \textit{i}-th
microcostituent of the composite system. The situation can be summarized as
follows:

\begin{itemize}
\item No one of the space-time points $R$ and/or $L$ at which an apparatus is
switched on belongs to the volume $\widetilde{V}(\sigma(P),\sigma_{0})$ lying
between $\sigma_{0}$ and $\sigma(P)$ (this last surface being the one
defined in section 6). Then the event ``the observable $\Theta^{(i)}$ is
indefinite'' is true.

\item If any one of the space-time points $R$ and/or $L$ at which an
apparatus is
switched on belongs to $\widetilde{V}(\sigma(P),\sigma_{0})$ then the
corrisponding
event is ``the microproperty related (or anticorrelated) to the outcome of
the isospin
component which has been measured'' is definite. The probability of its
value is
precisely determined by the theory, the actual occurence of one of the possible
outcomes is a genuinely random event.
\end{itemize}

Note that, in accordance with the above statements and when the apparatus
at $R$ is on,
the assertion ``the observable $\Theta^{(1)}$ of particle $1$ is
indefinite" holds
for all space-time points of the world line of  microsystem 1 preceeding
the point in
which the future light cone from $R$ intersects its world line. The event
``the property of microcostituent 1 is definite and it is opposite to the
outcome of the
measurement at $R$ on system 2'' emerges when system 1 reaches the future
light cone of
the measurement event.

Suppose now one is interested in the event characterizing a precise
space-time point of the world line of a macroscopic measuring apparatus. As
already remarked and as it should be evident by our argument the event for
such a system is always precisely defined and it corresponds to one of the
alternatives ``the pointer points to the $r$ (ready) position, it points to
+, it points to -''. It is important to stress that this holds for both
world lines of the apparata independently of the fact that they are switched
on or off (in which case they are always in the $r$ state) and independently
of the fact that only one or both of them are switched on. In spite of this
fact, in the case in which both apparatuses are on, the ``definite events''
referring to space-time points following, on their world lines, the
objective space-time points at which the system-apparatus interactions take
place are always opposite, i.e., the perfect anticorrelations characterizing
SQM predictions are respected.

\subsection{Counterfactuals and nonlocality}

We consider it appropriate to call the attention of the reader on the
extremely relevant implications of the nonlocal nature of quantum theory for
counterfactual arguments within a relativistic context. We recall, first of
all, that
we have related the possibility of making counterfactual assertions about
an objective
space-time point to the consideration of the past light cone from the
considered
point. This is unavoidable within a context like the present one  in which the
dynamics is fundamentally irreversible, so that the absolute
past plays a basic role for any consideration concerning the absolute
future.

In this subsection we want to analyze in greater details the  problems
which are specifically related to nolocality, to have the opportunity of
stressing some
subtle points. We start by considering a quite  simple
objection to our  way of dealing with counterfactual assertions which could be
raised by a naive reader: in the two particle case of subsections
7.4 and 7.6, why an observer who is on the world line of particle 2 at a point
{\it Q} in the immediate future of the point
$R$ at which the isospin component $T_{2z}$ of this particle has been
measured (and
found, e.g. to have the value --1) is not allowed to make the statement ``if an
apparatus were switched on on the world line of particle 1 at a point
\textit{L}
which is space-like with respect to both $R$ and $Q,$ such an apparatus would
register with certainity the outcome +1"? Here is where nonlocality enters.
To claim that the above statement is appropriate means to assume that the
accessibility sphere from the actual world is represented by all those
worlds in which the antecedent, i.e., the fact that the outcome
at \textit{R} has been --1, is true. If the theory were local, i.e. if
the outcome at a given point were totally independent from all what is
going on at space-like separations, then the argument would be
perfectly correct. But, as we know, this is not the case. To allow the
reader to
grasp this subtle point as well as the argument we will present below, we
invite him to
consider the three following situations and the related statements:

\begin{itemize}
\item  We consider an actual world in which everything is like in the two
particle
case of subsection 7.4 and, moreover, {\it both apparatuses are on}. We also
consider an observer along the world line of particle 2 immediately after
the point $R,$ who is aware of the outcome of the measurement and is also
aware of the fact that the apparatus at $L$ is on. Then he can argue in the
following way: I know that the apparatus $A$ (the one at $R$) has
registered ``the
isospin has the value --1'', I also know that another apparatus is on at
$L$ and that
the theory guarantees that the final outcomes are always anticorrelated. I can
then claim that the apparatus $B$ at $L$, at any instant following (on its
world
line) the one of its interaction with particle 1, registers the value +1.
\end{itemize}

Note that the above \textit{is not} a counterfactual argument since it makes
exclusive reference to the actual world. Moreover, as it should be clear
from our
analysis in the previous sections, such an argument is perfectly legitimate and
correct.

\begin{itemize}
\item  We consider now the same situation, but we assume
that, in the actual world, {\it the apparatus at $L$ is off} (note that
this is a
statement about a precise space-time event, meaning that the apparatus is
off when the particle might trigger it). Let us consider once more our observer
and his reasoning: I know that the apparatus $A$ at $R$ has registered ``the
isospin has the value --1'' and I also know that the theory guarantees that
if two
apparata were present and switched on at \textit{L}
and
\textit{R} their outcomes would be perfectly  anticorrelated. I can then
claim that
\textit{if} the apparatus at $L,$
\textit{were on,} at any instant following (on its world line) the one
of its interaction with particle 1, it would register the value +1.
\end{itemize}

This, as the reader has certainly clear, is a genuinely counterfactual
argument and, according to our criterion (but as we will show, completely in
general, due to the nonlocal character of the theory) is definitely
illegitimate.

\begin{itemize}
\item  The final case is exactly the same as the previous one, and, in
particular, {\it the apparatus at $L$ is switched off}. The only difference
consists in the statement made by the observer. Such statement does not refer
to a point which is space-like with respect to him, but to one, let us say $%
T,$ along the world line of particle 1  lying  in the future light cone
of the observer. The he can claim:  I know that the apparatus at $%
R$ has registered ``the isospin has the value --1'' and I also know that the
theory guarantees that the outcomes of pairs of apparata at \textit{R} and
\textit{T}  (which are time-like separated) are always anticorrelated. So,
in spite of
the fact that in the actual world no apparatus is on along the world line
of particle
1, I can claim (which, for time--like separations also means {\it predict})
that  if
such an apparatus were on at $T,$ then it would register the outcome
``the isospin equals $+1$''.
\end{itemize}

We stress that this is a genuinely counterfactual argument and we also
stress that,
according to our criterion and to standard wisdom, is a perfectly
legitimate one.

To further clarify why the position analyzed in the second of the above
cases is inappropriate and to show that the fundamental reason for this
derives from
the basically nonlocal nature of physical processes which has been so
appropriately
brought to the attention of the scientific community by the analysis of J.S.
Bell \cite{J.S. Bell}, we will consider now an hypothetical (and certainly
possible)
deterministic completion of our toy model, i.e. a
nonlocal deterministic hidden variable theory equivalent to it.  The theory
will
be characterized by hidden variables which can take values from a set
$\Lambda$ and
whose knowledge would unambiguously determine all outcomes of all conceivable
measurement processes.
\begin{figure}[htb]
\begin{center}
{\includegraphics[scale=0.4]{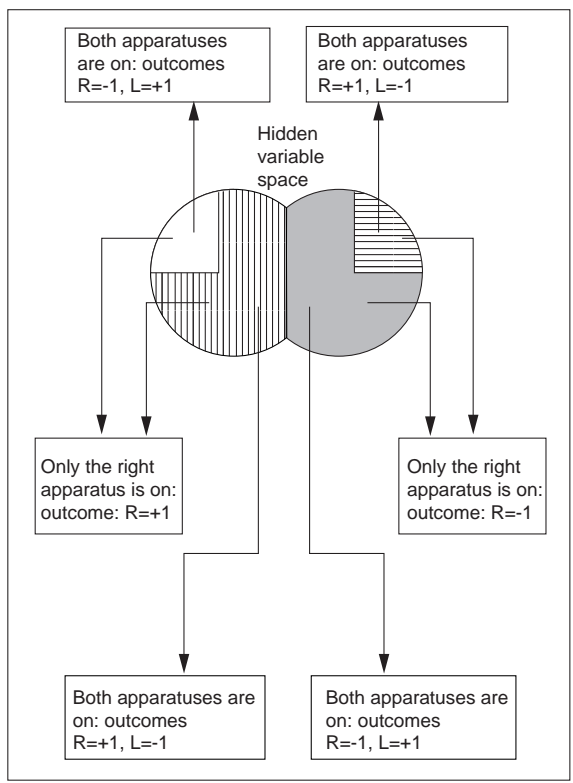}}
\caption{A diagram illustarting, with reference to an hypothetical
relativistic and
nonlocal hidden variable mode, the problems one has to face when resorting  to
counterfactual reasoning.}
\end{center}
\end{figure}
Within the set $\Lambda$ we identify (Fig.6) two subsets
$\Lambda _{1}(2,+)$ and
$\Lambda _{1}(2,-)$ such that in the case in which only the apparatus at
$A$ is switched on, if the actual value $\lambda $ of the hidden variable
belongs to $\Lambda _{1}(2,+)$ [$\Lambda _{1}(2,-)]$ then the outcome of the
measurement is $+1$ ($-1).$ Let us now consider the case in which both
apparatuses are switched on. Then the fact that locality is violated implies
that there exist a non empty subset $\Lambda _{1-2}$ of $\Lambda _{1}(2,+)$
such that, for $\lambda \in \Lambda _{1-2}$ the outcome at $A$ is --1 and the
one at $B$ is +1 \footnote {Actually nonlocality implies that this should
occur for at
least some of the pairs of perfectly correlated observables of the
constituents. For
simplicity we assume here that this actually happens for the pair we are
interested
in. This does not change in any way the conceptula implications of our
analysis.}.

To
judge of the validity of a counterfactual statement like ``if besides the
apparatus at
$R$ also the apparatus at
$L$ were on then the outcome at $L$ would be ...'' one has to identify the
accessibility sphere from the actual world. Which is the appropriate
criterion to
characterize the worlds which are nearest to the actual one? If one
claimed that they are those characterized by the same value of the hidden
variables, then for the subset $\lambda \in \Lambda _{1-2}$ of $\Lambda
_{1}(2,+)$ the appropriate specification which must replace the dots in the
previous sentence would be ``$+1$'' (while the outcome at $R$ should be
characterized by the value opposite to the one which occurs with certainty
when only such apparatus is on). On the contrary, if one took the
position that the worlds which are nearest to the actual one are those in
which the outcome at $R$ is the same as the one obtained in the actual
world, then one would be including alternative worlds characterized by a
value of the hidden variables belonging to an appropriate subset of $\Lambda
_{1}(2,-)$, i.e. worlds such that if only the apparatus at $R$ were on
would give the outcome opposite to the one which has occurred in the actual
world.

This elementary example should have made clear how delicate
is the handling of counterfactual statements in a relativistic nonlocal
context and why the prescription we have adopted is the only consistent and
logical
one. For a further
discussion of this delicate point we refer the reader to ref.\cite{GG}.

\section{Relativistic dynamical reduction and local measurements of nonlocal
observables}

In this section, with reference to our toy model, we will
reconsider the experimental set-up of section 3 devised to measure nonlocal
observables of a two particle system by resorting to local interactions and
detections. The first important remark is that the dynamical evolution has
to be enriched (with respect to the examples analyzed in the previous
section) to take into account the further interaction processes between the
microprobes ([3], [2*], [4*] and [6], [4*], [6*]) and the microscopic
constituents
((1) and (2)). Just because these interactions involve only
microscopic systems they do not give rise to reduction processes and are
accounted by
the linear and deterministic evolution summarized in eqs. (3.15)-(3.18).
Obviously, also such interactions occur at precise, objective space-time
points, so that the whole processs is perfectly covariant.

The first appropriate step is to analyze the evolution from the initial
space-like surface $\sigma _{0}$ to a surface $\sigma _{1}$ which has
''crossed'' a region in which  the interactions between the probes
and the particles occur (see Fig. 7, where the considered region  is the one
at right, lying on the world line of particle 2).

We still assume that the
initial state is the singlet isospin state. In order to evaluate the
evolution due to the interactions of particle 2 with the probe particles
[6], [4*] and [6*] we remark that such an evolution is governed by the
appropriate  part of the operator $U=\widetilde{U_{z}}U_{y}U_{z}$ defined
after eq.(3.13). Such an operator can be written as the direct product of
an operator acting on particle 2 and one acting on particle 1 (this
expresses the local nature of the interactions with the probes):
\begin{eqnarray}
& U=U^{(1)}\otimes U^{(2)},
\label{f1}
\\
& U^{(2)}=[P_{z+}^{(2)}P_{L}^{(6)}+P_{z-}^{(2)}P_{R}^{(6)}]\cdot %
[P_{y+}^{(2)}P_{L}^{(4^{*})}+P_{y-}^{(2)}P_{R}^{(4^{*})}]\cdot %
[P_{z+}^{(2)}P_{R}^{(6^{*})}+P_{z-}^{(2)}P_{L}^{(6^{*})}].
\nonumber
\end{eqnarray}
According to eq. (\ref{f1}), the effect of applying $U^{(2)}$ to the singlet
state is (as one proves by a rather involved calculation):
\begin{eqnarray}
& |\Psi (\sigma _{1})>\equiv U^{(2)}|Singlet>|\phi >
\label{f1b}
\\
&
=\frac{i}{2\sqrt{2}}\left( P_{R}^{(6)}|\phi >_{3,6}\right) \left[ \left(
P_{R}^{(4^{*})}-P_{L}^{(4^{*})}\right) |\phi >_{2^{*},4^{*}}\right] \left(
P_{L}^{(6^{*})}|\phi >_{3^{*},6^{*}}\right) |\uparrow _{1z},\uparrow _{2z}>
\nonumber \\
&
+\frac{1}{2\sqrt{2}}\left( P_{R}^{(6)}|\phi >_{3,6}\right) \left[ \left(
P_{R}^{(4^{*})}+P_{L}^{(4^{*})}\right) |\phi >_{2^{*},4^{*}}\right] \left(
P_{R}^{(6^{*})}|\phi >_{3^{*},6^{*}}\right) |\uparrow _{1z},\downarrow _{2z}>
\nonumber \\
&
-\frac{1}{2\sqrt{2}}\left( P_{L}^{(6)}|\phi >_{3,6}\right) \left[ \left(
P_{L}^{(4^{*})}+P_{R}^{(4^{*})}\right) |\phi >_{2^{*},4^{*}}\right] \left(
P_{L}^{(6^{*})}|\phi >_{3^{*},6^{*}}\right) |\downarrow _{1z},\uparrow _{2z}>
\nonumber \\
&
-\frac{i}{2\sqrt{2}}\left( P_{L}^{(6)}|\phi >_{3,6}\right) \left[ \left(
P_{L}^{(4^{*})}-P_{R}^{(4^{*})}\right) |\phi >_{2^{*},4^{*}}\right] \left(
P_{R}^{(6^{*})}|\phi >_{3^{*},6^{*}}\right) |\downarrow _{1z},\downarrow
_{2z}>.
\nonumber
\end{eqnarray}

\begin{figure}[htb]
\begin{center}
{\includegraphics[scale=0.5]{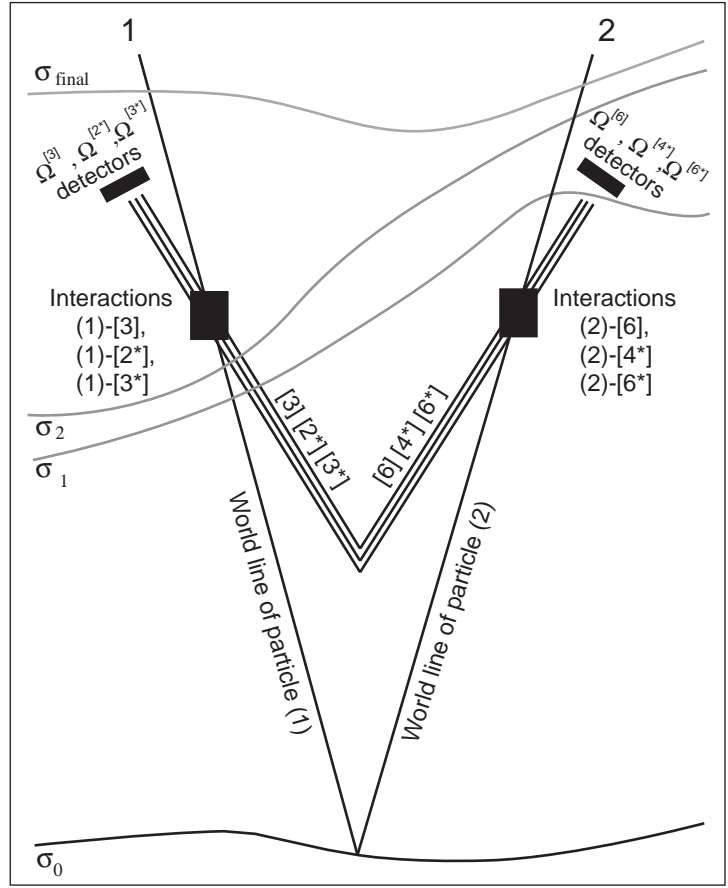}}
\caption{Possible space-like surfaces which is interesting to consider in
the case of a
nonlocal measurement of the square of the total isospin of a system of two
far-away particles.}
\end{center}
\end{figure}

Recalling the effect of the
operators $P_{L}^{(i)}$ and $P_{R}^{(i)}$ on the states $|\phi >_{r,i}:$%
\begin{equation}
P_{L}^{(i)}|\phi >_{r,i}=\frac{1}{\sqrt{3}}\left[
|0,-1>_{r,i}+|1,1>_{r,i}+|-1,0>_{r,i}\right]
\end{equation}
\[
P_{R}^{(i)}|\phi >_{r,i}=\frac{1}{\sqrt{3}}\left[
|0,1>_{r,i}+|1,0>_{r,i}+|-1,-1>_{r,i}\right]   \label{f3}
\]
we can rewrite the above state (\ref{f1b}) as:
\begin{eqnarray}
& | \Psi (\sigma _{1}) \rangle \equiv U^{(2)}|Singlet>|\phi >=
\label{f4}
\\
& \frac{1}{6\sqrt{6}}[i|A>|\uparrow _{1z},\uparrow _{2z}>
+|B>|\uparrow _{1z},\downarrow _{2z}>
-|C>|\downarrow _{1z},\uparrow _{2z}>-i|D>|\downarrow _{1z},\downarrow
_{2z}>],
\nonumber
\end{eqnarray}
where the probe states $|A>,$ $|B>,$ $|C>$ and $|D>$ are given by:
\begin{eqnarray}
|A>
& = &
\left( |0,1>_{3,6}+|1,0>_{3,6}+|-1,-1>_{3,6}\right)
\otimes
  \label{f5}
\\
& &
[|0,1>_{2^{*},4^{*}}+|1,0>_{2^{*},4^{*}}+|1,-1>_{2^{*},4^{*}}
-|0,-1>_{2^{*},4^{*}}-|1,1>_{2^{*},4^{*}}-|-1,0>_{2^{*},4^{*}}]
\otimes
\nonumber \\
& &
\left(
|0,-1>_{3^{*},6^{*}}+|1,1>_{3^{*},6^{*}}+|-1,0>_{3^{*},6^{*}}\right) ,
\nonumber
\end{eqnarray}

\begin{eqnarray}
|B>
& = &
\left( |0,1>_{3,6}+|1,0>_{3,6}+|-1,-1>_{3,6}\right)
\otimes
\label{f6}
\\
& &
[|0,1>_{2^{*},4^{*}}+|1,0>_{2^{*},4^{*}}+|1,-1>_{2^{*},4^{*}}
+|0,-1>_{2^{*},4^{*}}+|1,1>_{2^{*},4^{*}}+|-1,0>_{2^{*},4^{*}}]
\otimes
\nonumber \\
& &
\left(
|0,1>_{3^{*},6^{*}}+|1,0>_{3^{*},6^{*}}+|-1,-1>_{3^{*},6^{*}}\right) ,
\nonumber
\end{eqnarray}

\begin{eqnarray}
|C>
& = &
\left( |0,-1>_{3,6}+|1,1>_{3,6}+|-1,0>_{3,6}\right)
\otimes
\label{f7}
\\
& &
[|0,1>_{2^{*},4^{*}}+|1,0>_{2^{*},4^{*}}+|1,-1>_{2^{*},4^{*}}
+|0,-1>_{2^{*},4^{*}}+|1,1>_{2^{*},4^{*}}+|-1,0>_{2^{*},4^{*}}]
\otimes
\nonumber \\
& &
\left(
|0,-1>_{3^{*},6^{*}}+|1,1>_{3^{*},6^{*}}+|-1,0>_{3^{*},6^{*}}\right) ,
\nonumber
\end{eqnarray}

\begin{eqnarray}
|D>
& = &
\left( |0,-1>_{3,6}+|1,1>_{3,6}+|-1,0>_{3,6}\right)
\otimes
\label{f8}
\\
& & [|0,-1>_{2^{*},4^{*}}+|1,1>_{2^{*},4^{*}}+|-1,0>_{2^{*},4^{*}}
-|0,1>_{2^{*},4^{*}}-|1,0>_{2^{*},4^{*}}-|1,-1>_{2^{*},4^{*}}]
\otimes
\nonumber \\
& &
\left(
|0,1>_{3^{*},6^{*}}+|1,0>_{3^{*},6^{*}}+|-1,-1>_{3^{*},6^{*}}\right) .
\nonumber
\end{eqnarray}

We consider now the further evolution from $\sigma _{1}$ to $\sigma _{2}.$
In this process a micro-macro interaction between the probe particles and
the detectors  $\Omega ^{[6]},$ $\Omega ^{[4^{*}]}$ and $\Omega ^{[6^{*}]}$
takes place and  leads almost immediately
to precise locations of their macroscopic pointers. From the above equations
we see that all possible triplets of oucomes can occur. Just to fix our
mind, let us suppose that reduction has lead to the three outcomes $\Omega %
^{[6]}=0,$ $\Omega ^{[4^{*}]}=0$ and $\Omega ^{[6^{*}]}=0.$ The state after the
completion of this process is:
\begin{eqnarray}
& | \Psi (\sigma _{2})>
=
\label{f9}
\\
&
\frac{N}{6\sqrt{6}}\{i\left[ |1,0>_{3,6}\otimes \left(
|1,0>_{2^{*},4^{*}}-|-1,0>_{2^{*},4^{*}}\right) \otimes
|-1,0>_{3^{*},6^{*}}\right] \otimes |\uparrow _{1z},\uparrow _{2z}>
\nonumber \\
&
+\left[ |1,0>_{3,6}\otimes \left(
|1,0>_{2^{*},4^{*}}+|-1,0>_{2^{*},4^{*}}\right) \otimes
|-1,0>_{3^{*},6^{*}}\right] \otimes |\uparrow _{1z},\downarrow _{2z}>
\nonumber \\
&
-\left[ |-1,0>_{3,6}\otimes \left(
|1,0>_{2^{*},4^{*}}+|-1,0>_{2^{*},4^{*}}\right) \otimes
|-1,0>_{3^{*},6^{*}}\right] \otimes |\downarrow _{1z},\uparrow _{2z}>
\nonumber \\
&
-i\left[ |-1,0>_{3,6}\otimes \left(
|-1,0>_{2^{*},4^{*}}-|1,0>_{2^{*},4^{*}}\right) \otimes
|1,0>_{3^{*},6^{*}}\right] \otimes |\downarrow _{1z},\downarrow _{2z}>,
\nonumber
\end{eqnarray}
$N$ being a normalization factor.

And now comes the final step. We consider the subsequent evolution to a
final state which is associated to a space-like surface in the future of $%
\sigma _{2},$ obtained by crossing  the region of Fig. 7 where the
interactions of particle 1 with the probe particles takes place
(subsequently the
detectors at left will simply register the definite situation of the probes).
Such evolution is governed by the unitary operator
$U^{(1)}$ which has the same form as $U^{(2)}$ of eq. (\ref{f1}) with the
apices [3],
[2*] and [3*] replacing [6], [4*] and [6*]. With a rather tedious
calculations one gets
the result:
\begin{equation}
|\Psi (\sigma _{final})>=\left[ |0,0>_{3,6}\otimes |0,0>\otimes
|0,0>_{3^{*},6^{*}}\right] \otimes |Singlet>.  \label{f10}
\end{equation}
Obviously, one can perform the calculation by assuming any other possible
set of outcomes of the detectors at the right wing. For instance, if one
chooses the set  $\Omega ^{[6]}=1,$ $\Omega ^{[4^{*}]}=0$ and $\Omega %
^{[6^{*}]}=0$ and goes through precisely the same calculation, one gets, in
place of eq. (\ref{f10}) the following one:
\begin{equation}
|\Psi (\sigma _{final})>=\left[ |-1,1>_{3,6}\otimes |0,0>\otimes
|0,0>_{3^{*},6^{*}}\right] \otimes |Singlet>.  \label{f11}
\end{equation}
As expected, the final state of the system is always the singlet and the
states of the various pairs of apparata appearing in it are such that the sum
of their eigenvalues is always zero, as it must be in order that the
formalism be internally consistent. The same happens for
different initial states and for any possible choice of the outcomes at
the right wing. We believe that this simple exercise should have made clear
to the
reader the elegance and the logical coherence of the formalism.

Obviously, mastering
completely the theoretical scheme allows us to tackle, also in the more general
case of nonlocal measurements, the problem of property attribution to the
system.
Since such measurements involve a pair of
space-time points one has to identify the space-like hypersurface which
has to be considered for drawing the relevant conclusions. According to our
criterion, such a hypersurface is the
boundary of the two past light cones from the pair of points one is taking
into account on the world lines of the two particles. It is then obvious
that, if
consideration is given to a pair of points which preceed the two
interaction regions
(both at right and at left) of the particles with the probes, the theory
associates to
such a surface the singlet state and one can claim that ``the composite system
possesses the objective property of having total isospin equal to 0''. Such an
assertion is confirmed by the subsequent measurement procedure and the
communication
between the two obserbers concerning the outcomes they have registered. On the
contrary, if one considers a pair of points such that one follows and the
other one
preceeds the particle-probe interactions, i.e. two points like those in
which the
hypersurface $\sigma _{1}$ of Fig. 7 crosses the two world lines, the state
describing
the overall situation is the one of eq.(%
\ref{f1b}) for which the square of the total isospin is undefined. One can
go on reasoning in the just outlined way. We have no space to discuss all
aspects of
such a procedure in detail, we simply point out  that, with
the chosen criterion the unfolding of the process and the statements one is
led to
make are quite natural.

The last comment which is worth making concerns the other
possible (in principle) choice for the space-like surface one uses to
identify the
possessed properties. As already remarked, one could make reference to the
future
rather than to the past light cone without meeting any logical
contradiction. However, we cannot avoid stressing that the ensuing picture
is rather
peculiar. The reasons should be quite obvious: in order to speak of an
objective
situation at an objective space-time point it is much more natural to make
reference
to its absolute past than to the past plus the whole region which is
space-like with
respect to it. In particular, we point out that when the future light cone
criterion
is adopted, in the situation of section 7.4 and in the case in which only the
apparatus at
$R$ is switched on, the isospin of particle 1 becomes definite at the
space-time point
at which the past light cone from $R$ intersects its world line. In our
opinion, this
sort of (luminal) backward causation is less natural than the forward one
implied by
the alternative criterion. Even more peculiar aspects emerge when one takes
into
account nonlocal measurements. These reasons make our choice, whith its precise
physical implications, definitely preferable.

\section{Conclusions}

The detailed analysis we have performed should, in our
opinion, have made clear the conceptually most relevant features that any
relativistic model
of dynamical reduction must satisfy. It should also  have made clear that the
relativistic generalizations of CSL satisfy, in principle, all
requirements we have identified. As already mentioned, such models encounter
some technical difficulties. But we believe that the present analysis makes
plausible that the main physical ideas at the basis of the considered
approaches must be essential ingredients of any attempt to
account for the macro-objectification process  respecting the relativistic
 requirements and emboding the fundamental nonlocal aspects of
natural processes.

\section*{Acknowledgements}

We thank Dr. F. Camana for a careful checking of some of the cumbersome
calculations and Drs. A. Bassi and L. Marinatto for stimulating exchanges of
views and for help in preparing the LaTex version of the manuscript.

\end{document}